\documentclass[a4paper,12pt]{article}
\pdfoutput=1 
\usepackage{jheppub}
\usepackage{wrapfig}
\usepackage{color}
\usepackage{rotating}
\usepackage{epstopdf}
\usepackage{mathtools}

\title{High--energy limit\\ of collision--induced 
false vacuum decay}

\author{Sergei Demidov and}
\author{Dmitry Levkov}
\affiliation{Institute for Nuclear Research of the Russian Academy
  of Sciences,\\  60-th October Anniversary Prospect 7a, Moscow
  117312, Russia}
\emailAdd{demidov@ms2.inr.ac.ru}
\emailAdd{levkov@ms2.inr.ac.ru}


\abstract{We develop a consistent semiclassical description of
  field--theoretic collision--induced tunneling at arbitrary high
  collision energies. As a playground we consider a
  $(1+1)$--dimensional false vacuum decay initiated by a collision of
  $N$ particles at energy $E$, paying special attention to the realistic
  case of $N=2$ particles. We demonstrate that the cross section of
  this process is exponentially suppressed at all energies. Moreover,
  the respective suppressesion exponent $F_N(E)$ exhibits a
  specific behavior which is significant for our semiclassical method and assumed to
  be general: it decreases with energy, reaches absolute minimum $F =
  F_{min}(N)$ at a certain threshold energy $E = E_{rt}(N)$, and stays
  constant at higher energies. We show that the minimal suppression
  $F_{min}(N)$ and threshold energy can be evaluated using a special
  class of semiclassical solutions which describe exponentially suppressed
  transitions but nevertheless evolve in real time. Importantly, we
  argue that the cross section at energies above $E_{rt}(N)$ is
  computed perturbatively in the background of the latter solutions,
  and the terms of this perturbative expansion stay bounded in the
  infinite--energy limit. Transitions in the high--energy regime
  proceed via emission of many soft quanta with total energy $E_{rt}$;
  the energy excess $E-E_{rt}$ remains in the colliding particles till
  the end of the process.}
\begin{document} 
\begin{flushright}
INR--TH--2015--009
\end{flushright}
\vspace{-10.5mm}
\maketitle
\flushbottom

\section{Introduction}
\label{sec:intro}
Exponential suppression of probabilities precludes direct
observation of extraordinary tunneling phenomena such as
baryon number violation in instanton--like electroweak 
transitions~\cite{Belavin:1975fg,'tHooft:1976fv,Rubakov:1996vz} or 
spontaneous decay of allegedly false Higgs
vacuum~\cite{Krive:1976sg,Krasnikov:1978pu,Degrassi:2012ry,
  Buttazzo:2013uya, Bezrukov:2014ina}.
Quantum mechanical intuition suggests, however, that tunneling
probabilities grow with energy. Indeed, tunneling phenomena of the
above sort occur at higher rates~\cite{Ringwald:1989ee,
  Espinosa:1989qn} in two--particle collisions: 
\begin{equation}
\label{eq:1}
\sigma(E) \propto \mathrm{e}^{-F(E)/g^2}\;,
\end{equation}
where the suppression exponent\footnote{Dubbed ``holy grail
  function''~\cite{Mattis:1991bj}.} $F(E)$ decreases with collision
energy $E$, while $g$ is a small coupling constant. The central question
is whether the exponential suppression in Eq.~\eqref{eq:1} disappears
at sufficiently high energies and, if it does not, what is the value
of the suppression exponent at $E\to +\infty$.  

\begin{figure}
  \vspace{3mm}
  (a) \hspace{5.5cm} (b) \hspace{5cm} (c)

\vspace{-2.17mm}
\centerline{\begin{minipage}{5.42cm}
    \includegraphics[width=5.42cm]{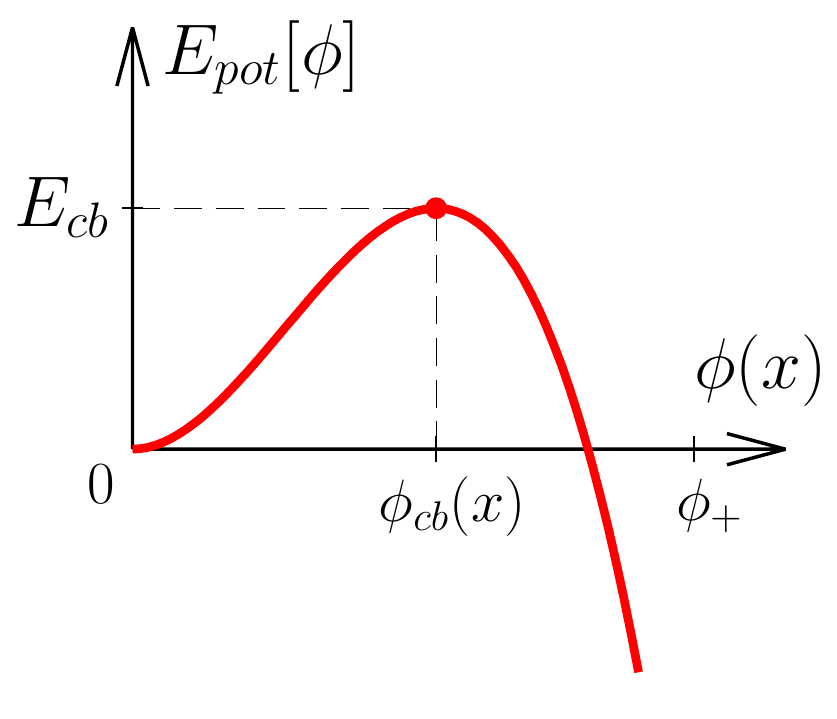}
  \end{minipage}\hspace{3mm}
  \begin{minipage}{4.33cm}
    \vspace{-0.975cm}
    \includegraphics[width=4.33cm]{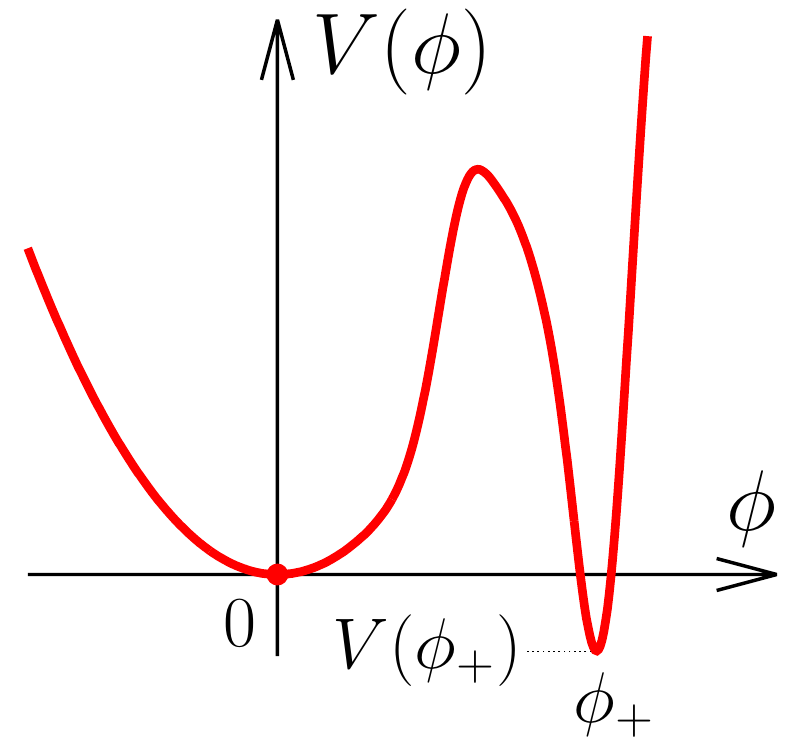}
  \end{minipage}\hspace{3mm}
  \begin{minipage}{6.5cm}
    \vspace{-5.42mm}
    \includegraphics[width=6.5cm]{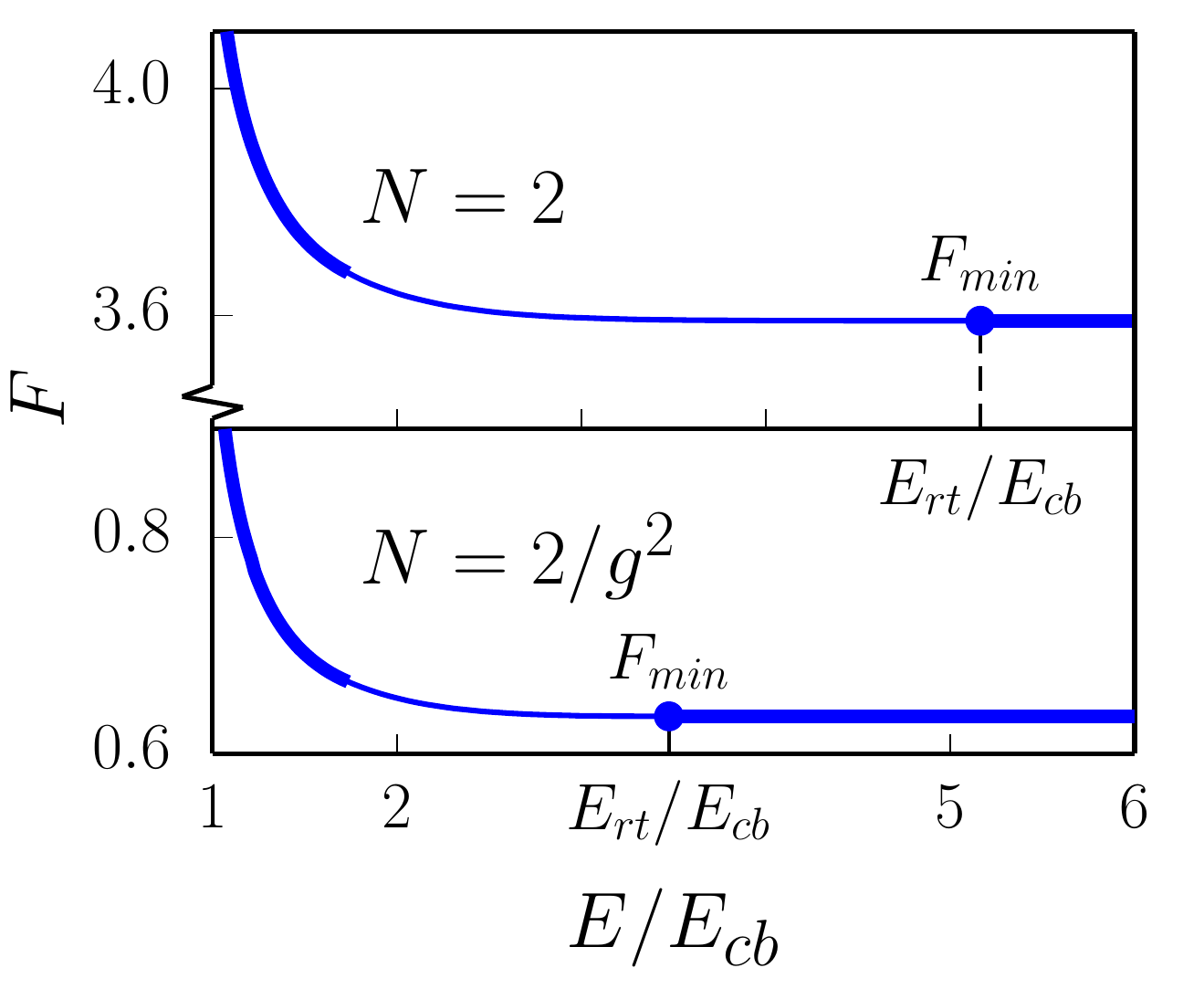}
   \end{minipage}} 
\caption{(a) A sketch of the potential barrier $E_{pot}[\phi]$
  between the vacua $\phi=0$ and $\phi = \phi_+$. (b)~Scalar potential
  $V(\phi)$. (c) Suppression exponent $F_N(E)$ at high
  energies. \label{fig:potential_energy}}
\end{figure}

This question is surprisingly nontrivial. Field--theoretic tunneling
involves barriers of finite heights $E_{cb}$ given by the energies of
the critical
bubble~\cite{Kobzarev:1974cp,Stone:1976qh,Coleman:1977py,Coleman:1978ae}
and sphaleron~\cite{Manton:1983nd,Klinkhamer:1984di}  in scalar and
gauge theories, see
Fig.~\ref{fig:potential_energy}a. Nevertheless, the respective
collision--induced transitions cannot become
unsuppressed\footnote{Gravitational interactions are
  completely different in this
  respect~\cite{Amati:2007ak, Dvali:2014ila}.} at
$E>E_{cb}$~\cite{Zakharov:1990xt,Zakharov:1991rp}. Consider 
e.g.\ massless fermions $\psi$ and $\bar{\psi}$ coupled with small Yukawa
constant $Y$ to the scalar sector of the theory.\footnote{It is
  straightforward to generalize this argument to other
  setups.} Their
contribution to the scalar self--energy $\Pi(Q^2)$ obeys 
dispersion relation~\cite{Zakharov:1991rp},
\begin{equation}
\label{eq:2}
\frac{d^2}{(dQ^2)^2}\Pi(Q^2) \Big|_{Q^2\to 0} = -\frac{8}{\pi Y^2} \int
\frac{dE}{E}\; \sigma_{tot}(E)\;,
\end{equation}
where $\sigma_{tot}(E)$ is the total annihilation cross 
section ${\psi   \bar{\psi} \to \mbox{anything}}$ at the
center--of--mass energy $E$ and we ignore irrelevant scalar
masses. If the fermion annihilation leads to  
unsuppressed over-barrier transitions at $E> E_{cb}$, the self--energy
$\Pi(Q^2)$ in Eq.~(\ref{eq:2}) receives large nonperturbative
contributions at small $Q^2$. This would contradict the standard
perturbation theory which is valid at low energies. Thus, the cross
section~(\ref{eq:1}) of collision--induced tunneling is
exponentially suppressed at~$E > E_{cb}$.

In this paper we describe collision--induced tunneling at high
energies. To be specific, we consider false vacuum decay in 
the $(1+1)$--dimensional scalar field model
\begin{equation}
\label{eq:8}
S[\phi] = \frac{1}{g^2} \int d^2 x \left[\frac12 (\partial_\mu \phi)^2 -
  V(\phi)\right]
\end{equation}
with false and true vacua at $\phi = 0$ and $\phi = \phi_+$,
respectively; the scalar potential $V(\phi)$ is 
shown in Fig.~\ref{fig:potential_energy}b. We work at weak coupling,
$g \ll 1$. At zero energy the vacuum $\phi=0$ decays 
spontaneously via formation of an expanding bubble with $\phi \approx
\phi_+$ inside. Below we study the same decay accompanied by a
collision of $N$ $\phi\mbox{--quanta}$ at energy $E$. We compute the
suppression exponent $F_N(E)$ of the corresponding inclusive cross
section at high energies.

Our numerical result\footnote{In numerical calculations we use a specific
  form of  $V(\phi)$ which is not significant at the moment.} for
$F_N(E)$ is presented in Fig.~\ref{fig:potential_energy}c. 
This function decreases with energy, reaches minimum
$F_{min}(N)$ at $E=E_{rt}(N)$ and stays constant at higher
energies. With the aid of the Rubakov--Son--Tinyakov
conjecture~\cite{Rubakov:1992ec} we 
extrapolate results to the two--particle initial state and 
find similar behavior of the exponent $F(E)\equiv F_{2}(E)$ in
Eq.~(\ref{eq:1})  (upper
panel in Fig.~\ref{fig:potential_energy}c). We conclude that
collision--induced false vacuum decay is exponentially suppressed at
arbitrary high energies. 

Energy--independent suppression exponent $F=F_{min}$ of collision--induced
tunneling at high energies was proposed in Ref.~\cite{Maggiore:1991vi}
and observed in toy models of Refs.~\cite{Voloshin:1993ks,
  Levkov:2004tf}. We find the same behavior in the full--fledged
field--theoretic model.

Besides, we demonstrate that induced false vacuum decay at the threshold $E
= E_{rt}(N)$, despite being exponentially suppressed, is described by
one--parametric family of complex semiclassical solutions $\phi =
\phi_{rt}(x)$ evolving in real time $t\equiv x^0$. These solutions
were introduced in Ref.~\cite{Levkov:2004ij} under the name  
{\it real--time instantons}. They satisfy complexified
classical field equations with certain boundary conditions in the
asymptotic past and future. The initial particle number $N$ parametrizing
the solutions enters the conditions at $t\to -\infty$. 
One can show~\cite{Levkov:2004ij} that the minimal suppression
$F_{min}(N)$ and respective energy $E_{rt}(N)$ are computed as
functionals on the real--time instantons. 

We argue on general grounds that if the real--time instantons 
exist for some collision--induced process, the respective
suppression exponent is energy--independent and equal to $F_{min}(N)$
at $E>E_{rt}(N)$. Thus, finding the family  of these solutions
in a given model, one obtains the exponent $F_N(E)$ in the entire
high--energy region $E> E_{rt}(N)$.

Most importantly, we demonstrate that the real--time instantons serve as 
backgrounds for the long--awaited~\cite{Maggiore:1991vi, Veneziano:1992rp}
perturbative description of collision--induced tunneling at high
energies, and the respective corrections are bounded in the
high--energy limit. This remarkable feature is in sharp contrast 
to the properties of perturbative
expansions in Euclidean backgrounds~\cite{Ringwald:1989ee,Espinosa:1989qn} which 
blow up~\cite{Mattis:1991bj,
  Tinyakov:1992dr}\cite{Rubakov:1996vz} at $E \sim E_{cb}$. Let us explain the difference
by considering scatterings of particles at energy $\Delta
E$ in different backgrounds. At a crude level the quantum particles
can be 
regarded as  small--amplitude high--frequency waves 
$\delta \phi \propto \mathrm{e}^{\pm i\Delta E t}$ added to the
background. In the Euclidean case $\delta \phi$ grows as 
$\mathrm{e}^{\Delta E \tau}$ with $\tau = -it$,  and nonlinear
backreaction effects become essential at high 
$\Delta E$. In other words, the perturbative expansion in $\delta \phi$
breaks down. In the opposite case of the real--time instanton the waves
$\delta \phi$ evolve adiabatically and do not change the 
soft background. Scattering of these waves can be described
perturbatively. 

Using the above observation, we propose a working perturbative
scheme~\cite{Rubakov-conjecture} for evaluating the inclusive cross
section of collision--induced tunneling as series in $g^2$ in the most
interesting case of high energies $E>E_{rt}(2)$ and two initial
particles. The receipt is as follows. One starts by developing a
formal perturbation theory in the background of a real--time
instanton $\phi_{rt}(x)$ with parameter $N=N_0$. Namely, one 
considers the $(n+2)$--point Green's function
\begin{equation}
\label{eq:36}
{\cal G}_{rt} \equiv \langle \Psi_{rt} | \phi(x_1) \dots \phi(x_{n+2}) |
\Psi_0\rangle  =
\int {\cal D} \phi \; \Psi_{rt}^{*}[\phi]\, \phi(x_1) \dots
\phi(x_{n+2})\, \mathrm{e}^{iS[\phi]} \,\Psi_0 [\phi]
\end{equation}
between the false vacuum $\Psi_0$ and the most probable final
state\footnote{A coherent
  state to be specified in the main body of the paper.} $\Psi_{rt}$ of
this real--time instanton. One 
substitutes
\[
\phi(x) \equiv \phi_{rt}(x) + g\delta \phi(x)
\]
into Eq.~(\ref{eq:36}) and evaluates the path integral over
fluctuations $\delta \phi$ as series in $g$. The terms in these
series depend on the auxiliary parameter $N_0$ of the 
real--time instanton which characterizes the background
configuration. Using the LSZ reduction formula, one extracts from the
Green's function the amplitude of the process $2\to n+\Psi_{rt}$,
where 
the final state contains $n$ particles on top of 
$\Psi_{rt}$. We argue that the
final states of this kind form a complete
set. Summing over them, one finds the inclusive cross
section of the 
collision--induced process. The final result is obtained in the limit
$N_0 \to 0$. 

Our method is different from the standard instanton perturbation
theory~\cite{Ringwald:1989ee, Espinosa:1989qn} in several
respects. First, we use the real--time instanton 
as a background. This guarantees stability of the perturbative
expansion at high energies. Second, we introduce an additional
parameter $N_0$ of $\phi_{rt}$ and send $N_0\to 0$ in the end of
calculations. Indeed, the real--time instanton with $N_0= 0$, as we
argue in the main body of the paper, is the saddle--point 
configuration for the path integral~(\ref{eq:36}). Formally it
describes transition from vacuum: two initial particles of the process
are represented by two $\phi$--factors in the integrand of
Eq.~(\ref{eq:36}) 
which do not affect the saddle--point solution. On the other hand, 
$\phi_{rt}(t,\, x)$ turns out to be singular at $N_0 = 0$.  We
therefore work with smooth saddle--point configurations at $N_0 > 0$
and recover\footnote{This makes our procedure a regularized 
  version of Landau method~\cite{Khlebnikov_Landau}~\cite{
    Voloshin:1993ks, Diakonov_Landau} of singular semiclassical
  solutions.} the  correct results in the limit $N_0 \to 0$. Third and
finally,  although below we evaluate only the leading inclusive
suppression exponent $F = F_{min}$ at $E>E_{rt}$, our perturbative
approach can be used for the prefactor and exclusive cross sections at
high energies.

\begin{wrapfigure}[9]{r}{4.8cm}

\vspace{-4mm}
\centerline{\hspace{-3mm}\includegraphics[width=5.2cm]{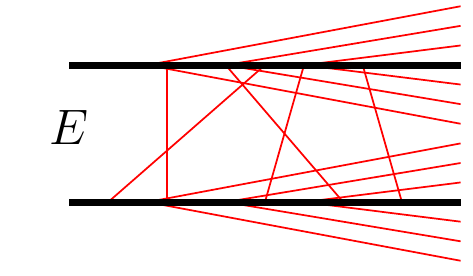}}

\vspace{-2mm}
\caption{Transitions at ${E>E_{rt}(2)}$.\label{fig:diagram_scheme}}
\end{wrapfigure}
With the help of the perturbative method we identify\footnote{For
  simplicity we consider the 
  process in a large finite volume.}
the dominant
mechanism of the collision--induced transitions at $E>E_{rt}(2)$ and
$N=2$, see Fig.~\ref{fig:diagram_scheme}. We observe that 
the colliding particles emit many soft quanta which form a bubble
of true vacuum with energy $E_{rt}(2)$; the energy excess
$E-E_{rt}(2)$  remains in the initial particles till the end 
of the process.  

From a general prospective our results support exponential
suppression of collision--induced tunneling at arbitrary high
energies and per se put on shaky ground proposed searches for
nonperturbative phenomena at future colliders~\cite{Ringwald:2003ns} 
or in cosmic ray events~\cite{Anchordoqui:2010hq, Anchordoqui:2011gy},
cf. Ref.~\cite{Acharya:2014nyr}. For example, it was found in 
Refs.~\cite{Bezrukov:2003qm, Bezrukov:2003er} that the suppression 
exponent of electroweak baryon number violation in two--particle collisions
is almost energy--independent at $E\sim 15 \;\mbox{TeV}$. If 
the minimum $E=E_{rt}(2)$ is somewhere near this point, the respective
cross section is suppressed at all energies by a deadly factor
$\mathrm{e}^{-F(15\, \;\mathrm{TeV})/\alpha_W} \sim 10^{-100}$, where
$\alpha_W$ is the electroweak coupling and we took numerics from
Refs.~\cite{Bezrukov:2003qm, Bezrukov:2003er}. To get reasonable
probabilities, one should 
consider models with tunneling rates raised by dynamical
mechanisms~\cite{Dutta:2008jt, Romanczukiewicz:2010eg, Lamm:2013ye,
  Papageorgakis:2014dma}, resort to strong
coupling~\cite{Morrissey:2005uza, Brandenburg:2006xu} or 
exotica~\cite{Dvali:2010jz}.

We argued in Ref.~\cite{Demidov:2011dk} that the 
collision--induced false vacuum decay in  $(1+1)$ dimensions turns
into production of kink--like soliton pairs from particles once
the  energy densities of the two vacua are leveled, $V(0) - V(\phi_+)
\to 0$. One expects that the properties indicated above hold in this
limit. In particular, the cross section of creating a pair of solitons
from two particles is exponentially small at all energies,
and the suppression exponent of this process does not depend on energy
above a certain threshold.

This paper is organized as follows. In Sec.~\ref{sec:grow-ampl-backgr}
we recall perturbative expansion in the background of a Euclidean
bounce. This technique reproduces   
exponentially growing collision--induced cross section 
at low energies but breaks down at $E\gtrsim E_{cb}$. We proceed
with moderate energies in Sec.~\ref{sec:from-euclidean-real} and
demonstrate how the bounce at $E\ll E_{cb}$ is connected with the
real--time instantons at $E= E_{rt}(N)> E_{cb}$. The latter
semiclassical solutions and transitions at $E>E_{rt}(N)$ are
considered in Secs.~\ref{sec:real-time-instatons} 
and \ref{sec:transitions-at-ee_rt}, respectively. In
Sec.~\ref{sec:pert-expans-at} we show that the same term of
perturbative expansion which led to  exponential growth of the cross
section with energy in perturbation theory around the bounce, gives
subdominant and exponentially decreasing contribution in the background of the
real--time instanton. We propose a working perturbative description of
collision--induced tunneling at $N=2$ and high energies in 
Sec.~\ref{sec:pert-descr-at}. Our results are summarized in
Sec.~\ref{sec:conclusions}.   

\section{Perturbative expansion in the background of a bounce}
\label{sec:grow-ampl-backgr}
Let us explain the difficulties with the collision--induced tunneling by
reviewing its low--energy description~\cite{Ringwald:1989ee,
  Espinosa:1989qn} in the model~(\ref{eq:8}). We will also
introduce terminology and sharpen contrast with the high--energy transitions.

\begin{wrapfigure}[18]{r}{5.0cm}
\vspace{-4mm}
\centerline{\hspace{4mm}\includegraphics[width=4.8cm]{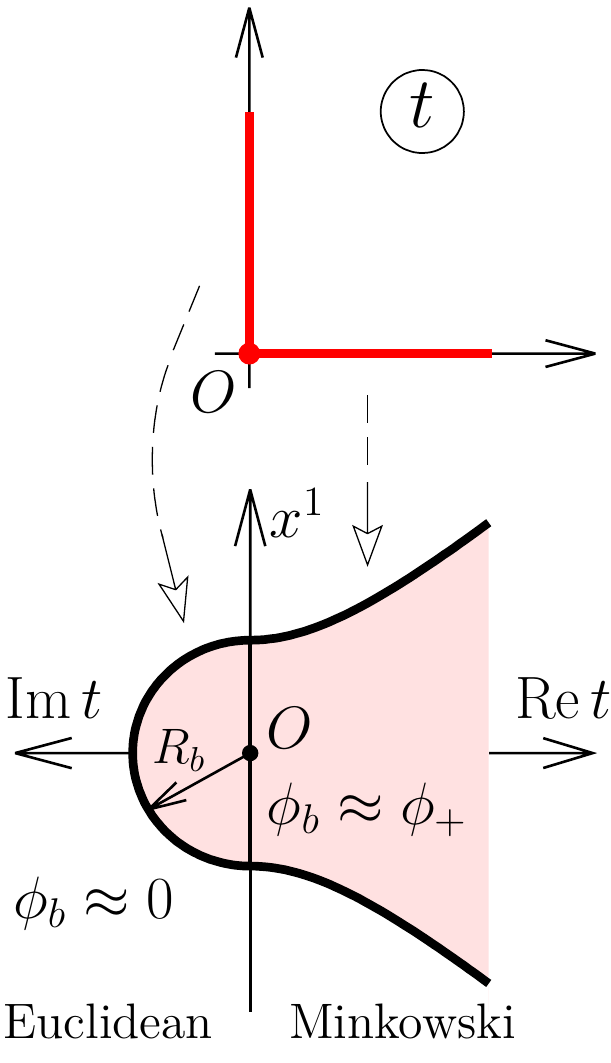}}

\vspace{-2.5mm}
\caption{Bounce $\phi_b(x)$. \label{fig:bounce}}
\end{wrapfigure} 
Spontaneous decay of false vacuum at $E=0$ is described by the celebrated
bounce solution~\cite{Coleman:1977py,Coleman:1978ae}  $\phi_b(x)$, see
Fig.~\ref{fig:bounce}. The latter has Euclidean and Minkowski parts 
representing nucleation of a true vacuum bubble and its expansion to
infinite size. Note that the bounce is Lorentz--invariant i.e.\ depends
on $x^2 \equiv x_\mu x^\mu$. At large negative $x^2$ it satisfies the
Klein--Gordon equation in the false vacuum and therefore behaves as
\begin{equation}
\label{eq:5}
\phi_b(x) \to  \frac{c_b}{2\pi}\, K_0 (m\sqrt{-x^2}) \qquad \mbox{as}
\qquad x^2 \to -\infty\;.
\end{equation}
Here $m$ is the mass in the false vacuum $\phi=0$. In
Eq.~(\ref{eq:5}) and below we exploit universal complex time $t\equiv
x^0$ which is real and pure imaginary in the respective parts of
the contour in Fig.~\ref{fig:bounce}. Parameter $c_b$ is related to
the bubble size $R_b$:  $\phi_b$ is of order $1$ at $x\sim R_b$. In
the thin--wall limit $m R_b \gg 1$ one obtains  $c_b \propto
\mathrm{e}^{mR_b}$, where the asymptotics of the Bessel function in
Eq.~(\ref{eq:5}) was used. 

To compute the amplitude of collision--induced tunneling, we consider
the Green's function 
\begin{align}
{\cal G}_{b}& \notag\equiv \langle \Psi_b|\phi(x_1)\dots
\phi(x_{n+2}) |\Psi_0\rangle \\& = \int {\cal D} \phi \; \Psi_b^*[\phi]\;
\phi(x_1) \dots \phi(x_{n+2}) \; \mathrm{e}^{iS[\phi]}\, \Psi_0[\phi]\;,
\label{eq:4}
\end{align}
where $|\Psi_0\rangle$ and $|\Psi_b\rangle$ are the false vacuum and
the final state of its decay at $E=0$, respectively. The latter describes expanding bubbles. 

The idea~\cite{Ringwald:1989ee,Espinosa:1989qn} is to evaluate 
the path integral in Eq.~(\ref{eq:4}) in the saddle--point
approximation treating $\phi(x_1) \dots \phi(x_{n+2})$ as a
prefactor. The relevant saddle--point configuration is the bounce
$\phi_b$. In the leading order one obtains, 
\begin{equation}
\label{eq:3}
{\cal G}_b = {\cal A}_b\int d^2 x_0\, 
\phi_b(x_1-x_0) \dots \phi_b(x_{n+2}-x_0)\,,
\end{equation}
where we introduced the bounce position $x_0$, ignored irrelevant saddle--point
determinant and denoted the bounce amplitude 
by ${\cal A}_b =  \Psi_b^*[\phi_b]\, \mathrm{e}^{iS[\phi_b]}\,
\Psi_0[\phi_b]$. Note that the
action $S$ and zero--energy states $\Psi_0$, $\Psi_b$ are
translation--invariant\footnote{Here and below we assume asymptotic limits
  $t_{i,f} \to \mp\infty$ when all quantities oscillating with the initial $t_i$
  or final $t_f$ times can be dropped. }  and therefore independent of
$x_0$.  The bounce amplitude is computed in
Refs.~\cite{Coleman:1977py,Coleman:1978ae}: with exponential 
precision 
\begin{equation} 
\label{eq:37}
|{\cal A}_b|^2 = \mathrm{e}^{-2\mathrm{Im}\, S[\phi_b]}\;,
\end{equation}
in particular, $|\Psi_b[\phi_b]|, \; |\Psi_0[\phi_b]| \sim 1$. We
remark that corrections to Eq.~(\ref{eq:3}) can be evaluated
perturbatively using the bounce as a background: one substitutes $\phi
= \phi_b + g\delta \phi$ into the path integral and calculates it as
series in $g^2$. 

Importantly, the large--$x$ asymptotics
of the bounce in Eq.~(\ref{eq:5}) reproduces correct residual at $k^2
= m^2$ of its Fourier transform, 
\begin{equation}
\label{eq:6}
\phi_b(k) \equiv \int d^2 x \, \mathrm{e}^{ik\cdot x} \phi_b(x-x_0)  =
\frac{ i c_b\, \mathrm{e}^{ik\cdot x_0} }{k^2 - m^2 + i\epsilon} + \mbox{regular part}\;,
\end{equation}
where we recalled that $K_0$ in Eq.~(\ref{eq:5}) is the Feynman
propagator in two dimensions; we denote $k\cdot x \equiv k_\mu 
x^\mu$ and assume $\epsilon \to +0$. The residue $c_b$ in
Eq.~(\ref{eq:6}) is $k$--independent or
``point--like''~\cite{Rubakov:1996vz}. This property is specific to
Euclidean solutions, as we argue below.  One immediately obtains the $2\to
n$ transition amplitude from the LSZ formula, 
\begin{equation}
\label{eq:7}
 {\cal A}_{2\to {n}} =  {\cal A}_b\,
\left(\frac{c_b}{g}\right)^{n+2}\;.
\end{equation}
To derive this expression, we Fourier--transformed Eq.~(\ref{eq:3}),
extracted the on--shell residues (\ref{eq:6}) and collapsed the
integral over $x_0$ into 
the $\delta$--function\footnote{It is absorbed in the phase space
  volume, as usual.} representing the energy--momentum conservation. Factors
$g$ in Eq.~(\ref{eq:7}) compensate for non--canonical normalization of
kinetic term in Eq.~\eqref{eq:8}. 

The amplitude (\ref{eq:7}) exponentially grows with energy $E$ if the
most probable final state contains ${n \approx E/m}$ nonrelativistic
particles. To confirm this guess about the most probable state, we derive the
inclusive cross section in Appendix~\ref{sec:mult-phase-volume}, 
\begin{equation}
\label{eq:10}
\sigma(E) = \sum_n \int |{\cal A}_{2\to n}|\, d\Pi_n 
= |{\cal A}_b|^2 \int
d^2 \lambda \, \mathrm{e}^{ iP\cdot  \lambda + \frac{|c_b|^2}{2\pi g^2} K_0(m\sqrt{-\lambda^2 +
  i\epsilon \lambda^0})}
\end{equation}
where the prefactors are ignored, $\Pi_n$ is the
$n$--particle phase space volume and $P^\mu = (E,0)$ is the total
momentum in the  center--of--mass frame. The variable $\lambda^\mu$ in
Eq.~(\ref{eq:10}) can be regarded as a typical 
Compton wavelength of the final particles: the latter become
nonrelativistic  at $\lambda\gg m^{-1}$. At $g\ll 1$ the integral in
Eq.~(\ref{eq:10}) is evaluated in the saddle--point approximation. The
extremum of the exponent is achieved at 
\begin{equation}
\label{eq:12}
\lambda^\mu_s = (-2i T,\, 0)\;,  \qquad\qquad  T = \frac{1}{2m}
\log \left(\frac{|c_b|^2 \sqrt{m}}{2g^2 E \sqrt{4\pi
    T}}\right)\;,
\end{equation}
where the asymptotics of the Bessel function was used.
This gives
\begin{equation}
\label{eq:38}
\sigma(E) = 
\mathrm{e}^{(2T+m^{-1})E - 2\mathrm{Im}\, S[\phi_b]}\;,
\end{equation}
see Eq.~(\ref{eq:37}).
In the thin--wall limit $V(\phi_+)\to 0$ we substitute $c_b\propto
\mathrm{e}^{mR_b}$ and obtain $T \approx R_b$ with corrections
proportional to $\log(mR_b)$. The cross section~(\ref{eq:38}) in this limit
coincides with that in Refs.~\cite{Kiselev:1992hk, Rubakov:1992gi},
see also Refs.~\cite{Voloshin:1985id, Selivanov:1985vt,
  Voloshin:1986zq}. 

The result~(\ref{eq:38}) demonstrates exponential growth of the cross
section with energy. It involves nonrelativistic final particles
when $\lambda_s \sim T$ is large i.e.\ at $E \ll m\, 
\mathrm{e}^{2mR_b}/g^2$, where the exponent comes from
$c_b$. Moreover, the relativistic regime is never reached 
because at $E\gtrsim  m/g^2$  weak coupling expansion becomes
unreliable. Indeed, relative corrections to the leading--order
result~(\ref{eq:3}) are estimated\footnote{Sophisticated  
  resummation~\cite{Arnold:1990va,Khlebnikov:1990ue} shows that the true
  expansion parameter is $g^2 n \ll 1$.}~\cite{Rubakov:1996vz} as 
$g^2 n^2$, where $n^2$ comes from combinatorics; they are already
large at $n \sim E/m \sim 1/g^2$. One concludes that correct description of 
collision--induced tunneling at $E \gtrsim E_{cb} \sim m/g^2$ should
incorporate backreaction of the final--state particles on
the semiclassical solution; we will pursue this approach in the next
Section. 

Let us point at two specific features of the low--energy
calculation. First, the bounce residue $c_b$ in Eq.~(\ref{eq:6})
does not depend on $k$ as $k \to +\infty$. Second, the final state
$\Psi_b$ has zero 
energy and factorizes in Eq.~(\ref{eq:3}). We will see that these two
properties do not hold for perturbative expansion about the
relevant high--energy solutions.

\section{From Euclidean to real--time solutions}
\label{sec:from-euclidean-real}
We demonstrated that collision--induced transitions  are no longer
described by the bounce at $E\gtrsim E_{cb}$. Since our interest lies in
high energies, we set this background aside and search for true
semiclassical solutions describing the false vacuum decay
in the  $N$--particle collisions. Consider the inclusive cross
section, 
\begin{equation}
\label{eq:13}
\sigma_N(E) = \sum_{\Psi_i,\, \Psi_f} \left| \langle
\Psi_f|\hat{U}(t_f,\, t_i)|\Psi_i; \, E,\,
N\rangle\right|^2 \approx \mathrm{e}^{-F_N(E)/g^2}\;,
\end{equation}
where $\hat{U}$ is the evolution operator, $t_{i,f} \to \mp \infty$,
we ignored the initial flux in the prefactor and introduced the
suppression exponent $F_N(E)$ in the approximate equality. The sum in
Eq.~(\ref{eq:13}) runs over all initial states 
$\Psi_i$ 
with energy $E$ and multiplicity $N$ in the false vacuum and final states 
$\Psi_f$ containing a bubble of true vacuum. Importantly, 
$\sigma_N(E)$ coincides at $N=2$ with the two--particle
collision--induced cross section $\sigma(E)$ and can be computed
semiclassically at $N \gg 1$. Moreover, Rubakov--Son--Tinyakov
conjecture~\cite{Rubakov:1992ec} states that the suppression exponent
$F_N(E)$ does not depend on $N$ at $N \ll 1/g^2$, see
Refs.~\cite{Tinyakov:1991fn, Mueller:1992sc, Bonini:1999kj,
  Levkov:2008} for confirmations. This means that the two--particle
exponent $F(E)$ in Eq.~(\ref{eq:1}) is obtained by extrapolating the
semiclassical result for $F_N(E)$ to $g^2 N \to 0$.

\begin{figure}[t]
\vspace{0.7cm}
\hspace{1.8cm}(a) \hspace{5.7cm} (b) \hspace{5cm} (c)

\vspace{-1.5cm}
\centerline{
\hspace{-12mm}\begin{minipage}{5.5cm}
\includegraphics[width=6.6cm]{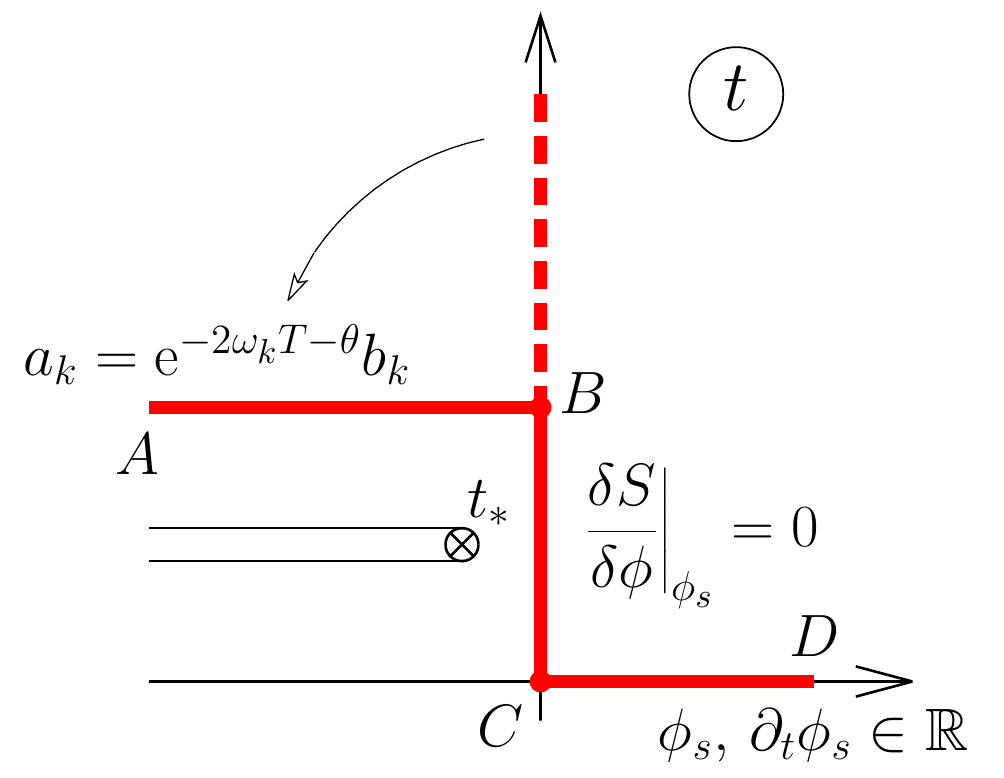}
\vspace{-3mm}
\end{minipage}\hspace{4mm}
\begin{minipage}{4.5cm}
\vspace{-5mm}
\hspace{7mm}
\begin{tabular}{|l|}
\hline
$\phi_+ \approx 2.56$\\
$m \approx 1.00$\\
$g^2 E_{cb} \approx 6.11$\\
$g^2\mathrm{Im}\, S[\phi_ b] \approx 86.7$ \\ 
$c_b \approx 7.4 \cdot 10^{4}$\\
\hline
\end{tabular}
\vspace{-4mm}
\end{minipage}\hspace{5mm}
\begin{minipage}{4.5cm}
\vspace{6mm}
\includegraphics[width=5.5cm]{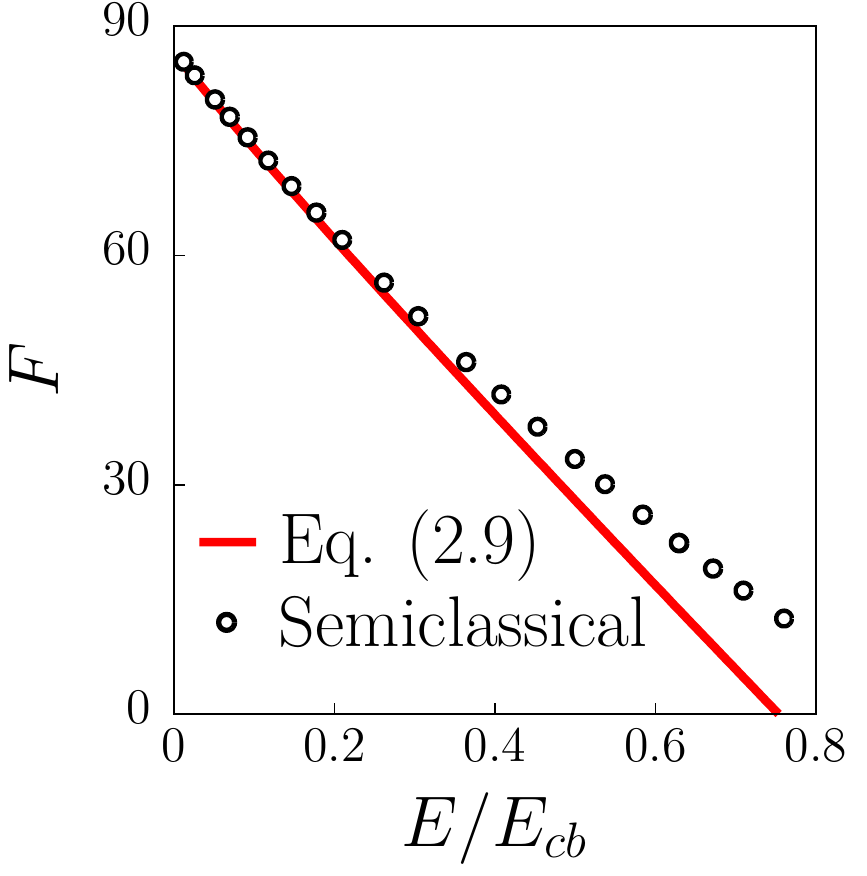}\\
\end{minipage}}

\vspace{-7mm}
\caption{(a) Contour in complex time for $\phi_s(t,\, x)$. Saddle--point
  equations and boundary conditions at $t\to \pm \infty$ are written
  near the respective parts of the contour. The nearest singularity
  $t_*$ of the solution is marked by the crossed circle attached to
  the branch cut (thin double line). 
(b) Physical quantities in the model~(\ref{eq:9}).
  (c)~Suppression exponent at low energies.
\label{fig:theta}}
\end{figure}

In direct semiclassical approach one writes a path integral for
$\sigma_N(E)$ and evaluates it at $g \ll 1$ using the saddle--point
configuration $\phi_s(t,\, x)$. In general, this configuration is
complex. The saddle--point equations for $\phi_s$ are
derived in Ref.~\cite{Rubakov:1992ec}, see the summary in
Fig.~\ref{fig:theta}a. Like the bounce, this configuration satisfies
the classical field equations\footnote{At the spatial boundary we
  impose the standard energy--conserving condition $\partial_x \phi_s
  \to 0$ at $x \to \pm\infty$.} $\delta S / \delta \phi = 0$ or 
\begin{subequations}
\label{eq:54}
\begin{equation}
\label{eq:56}
(\partial_t^2 - \partial_x^2) \phi_s  = -V'(\phi_s)
\end{equation}
 along the complex time contour $ABCD$ in Fig.~\ref{fig:theta}a. It 
contains an expanding bubble in the asymptotic future where the
solution is real, 
\begin{equation}
\label{eq:57}
\mathrm{Im}\, \phi_s, \; \mathrm{Im}\, \partial_t \phi_s \to 0 \qquad
\mbox{as} \qquad t\to +\infty\;.
\end{equation}
Peculiarities of $\phi_s(x)$ are related 
to the nontrivial initial state in Eq.~(\ref{eq:13}). 
Along the part $AB$ of the contour the saddle--point solution
describes motion of the particles prior to the collision: in the
asymptotic past it reduces to free waves
\begin{equation}
\label{eq:17}
\phi_s \to \int \frac{dk}{4\pi\omega_k} \left( a_k
  \mathrm{e}^{-i k \cdot x} + b_k^* \mathrm{e}^{i k \cdot
    x}\right)\qquad \mbox{as} \qquad t \equiv x^0\to -\infty\;, 
\end{equation}
where $k^\mu = (\omega_k,\, k)$, $\omega_k^2 = k^2 + m^2$, and the
limit is taken along the time contour of Fig.~\ref{fig:theta}a. In
Eq.~(\ref{eq:17}) we introduced the classical counterparts $a_k$, $b_k^*$
of the annihilation and creation operators with relativistic
normalization. They are related by the initial condition
\begin{equation}
\label{eq:39}
a_k = {\rm e}^{-2\omega_k T - \theta}b_k
\end{equation}
\end{subequations}
involving two Lagrange multipliers $T$ and
$\theta$ due to fixation of energy $E$ and the number $N$ of
colliding particles. The latter quantities are given by the standard
expressions 
\begin{equation}
\label{eq:14}
g^2 E= \int dk \, \frac{a_k b_k^*}{4\pi} \;, \qquad \qquad g^2 N = \int
dk \, \frac{a_k b_k^*}{4\pi \omega_k} \;.
\end{equation}
In the limit $T \to +\infty$ Eq.~(\ref{eq:39}) reduces to
vacuum condition $a_k = 0$ which corresponds to $E=N=0$. In this case $\phi_s(x)$ 
coincides with  the bounce solution $\phi_b(x)$. At finite $T$ and
$\theta$ the saddle--point solution describes transition at nonzero
$E$ and $N$. In
what follows we solve equations (\ref{eq:54}) and relate 
$(T,\, \theta)$ to $(E,\, N)$ by Eq.~(\ref{eq:14}).

Given the saddle--point configuration $\phi_s(x)$, one evaluates the
suppression exponent~\cite{Rubakov:1992ec} 
\begin{equation}
\label{eq:15}
F_N(E) = g^2(2\mathrm{Im}\, S[\phi_s] - 2ET - N\theta )
+ \mathrm{Im}\int dx \, \phi_s \partial_t \phi_s \Big|_{t = t_i}\;,
\end{equation}
where the last three terms are the initial--state
contributions. Importantly, the method of Lagrange multipliers implies
Legendre transform
\begin{equation}
\label{eq:16}
\partial_E F_N(E) = - 2g^2 T \;, \qquad\qquad \partial_N F_N(E) = -g^2
\theta\;, 
\end{equation}
which demonstrates that $T$ and $\theta$ are proportional to the derivatives of $F_N(E)$.

We solve the boundary value problem~\eqref{eq:54} for different $T$
and $\theta$
numerically. To this end we specify the scalar potential in
dimensionless units,
\begin{equation}
\label{eq:9}
V(\phi) = \frac{\phi^2}{2} \left[ 1 -
  vW\left(\frac{\phi-2}{u}\right)\right]\;, 
\end{equation} 
where $W(x)= \mathrm{e}^{-x^2} (x + x^3 + x^5)$, $u=0.4$, and the
energy density $V(\phi_+)=-0.4$ of the true vacuum is set by tuning
$v\approx 0.84$. Function (\ref{eq:9}) is plotted in
Fig.~\ref{fig:potential_energy}b. It is almost quadratic at $\phi
< \phi_+/2$ and nontrivial at larger $\phi$, so that 
waves in Eq.~(\ref{eq:17}) remain linear\footnote{Long 
  nonlinear evolution in other models is costly for numerical computations.} almost
up to their collision point.   

We numerically computed the physical quantities for the
potential (\ref{eq:9}),
see Fig.~\ref{fig:theta}b. To this end we have found the bounce
$\phi_b(t,\, x)$ and the critical bubble $\phi_{cb}(x)$, see
Ref.~\cite{Coleman:1978ae} for details. Recall that the bounce action
$2\mathrm{Im}\, S[\phi_b]$ in Eq.~(\ref{eq:37}) is the suppression
exponent of false vacuum decay at zero energy, whereas the energy
$E_{cb}$ of the critical bubble gives the
height of the potential 
 barrier between the vacua. We also extracted the bounce residue $c_b$
 from the asymptotics of $\phi_b(t,\, x)$ at $x_\mu x^\mu \to -\infty$, see
Eq.~(\ref{eq:5}). We remind that the coupling constant $g\ll 1$
scales out in the semiclassical calculations, cf. Eq.~(\ref{eq:8}).

We discretize equations \eqref{eq:54}
and introduce
uniform $N_t \times 2N_x$ lattice with sites $t_i$ and $x_j$
covering the contour $ABCD$ and space interval\footnote{
  Larger intervals are used at low energies due to larger sizes of the 
  respective solutions.} $(-L,\, L)$, $L =
7$; the spatial lattice spacing is $\Delta x\equiv L/N_x$. We use the
second--order  finite--difference approximation for the field
equation and trade Fourier transform in Eq.~(\ref{eq:17}) for its
discrete version. This turns the semiclassical boundary 
value problem into a set of $N_t \times 2N_x$ nonlinear algebraic
equations\footnote{The solutions are $P$--symmetric, $\phi_s(t,\, x) =
  \phi_s(t,\,-x)$, so we use only a half of the lattice with $x_j>0$.} for
$\phi_{ij} =\phi_s(t_i,\, x_j)$ which are solved by the Newton--Raphson
method~\cite{Press}. Detailed description of our numerical technique 
will be presented elsewhere~\cite{Demidov:2015}, see
Refs.~\cite{Kuznetsov:1997az}\cite{Bezrukov:2003er} for related 
works. In the subsequent Sections we will concentrate on numerical
solutions at high energies and small multiplicities. We will need 
large $N_t$ and $N_x$ because the typical frequencies ${\omega_k \sim
  E/N}$ of these solutions are high. In
particular, lattices $N_t\times N_x = 3000\times 150$ and $11000 \times
4000$ are required to reach acceptable numerical precision at $E\sim
E_{cb}$ and the highest energies, respectively. 

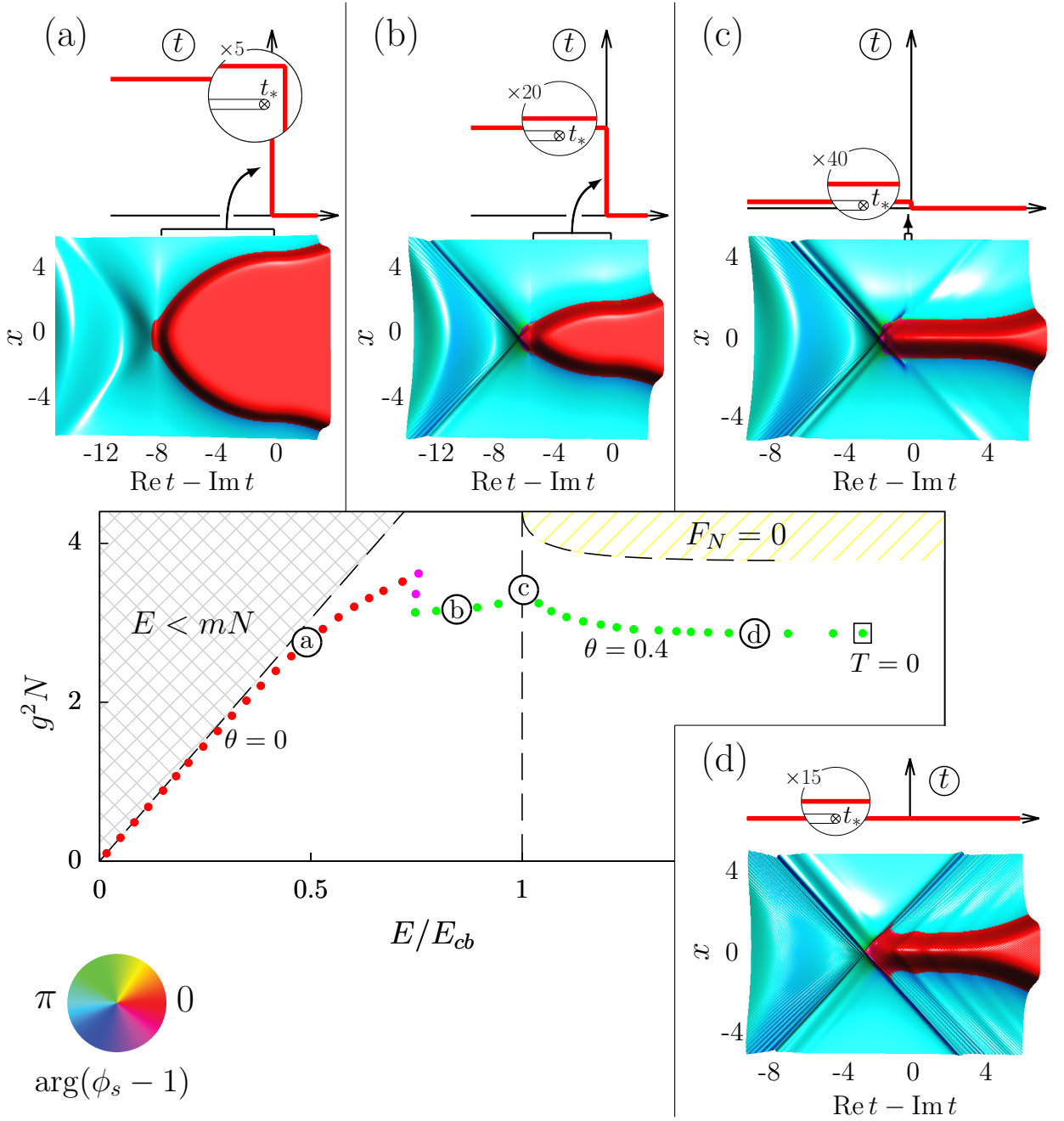
\begin{figure}[h!]
\vspace{-5mm}

\unitlength=0.00599999999999994\textwidth
\centerline{\begin{picture}(150,170.73)(-1,5)
\put(2,30){\includegraphics[width=142.5\unitlength]{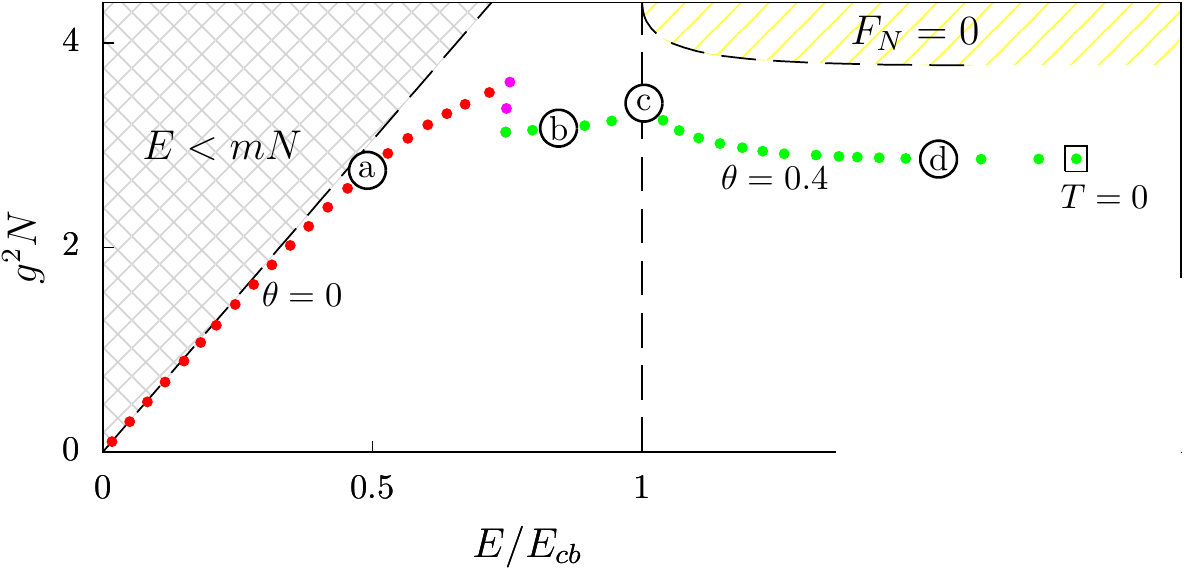}}
\put(7.5,142){\includegraphics[width=35.0\unitlength,angle=-90]{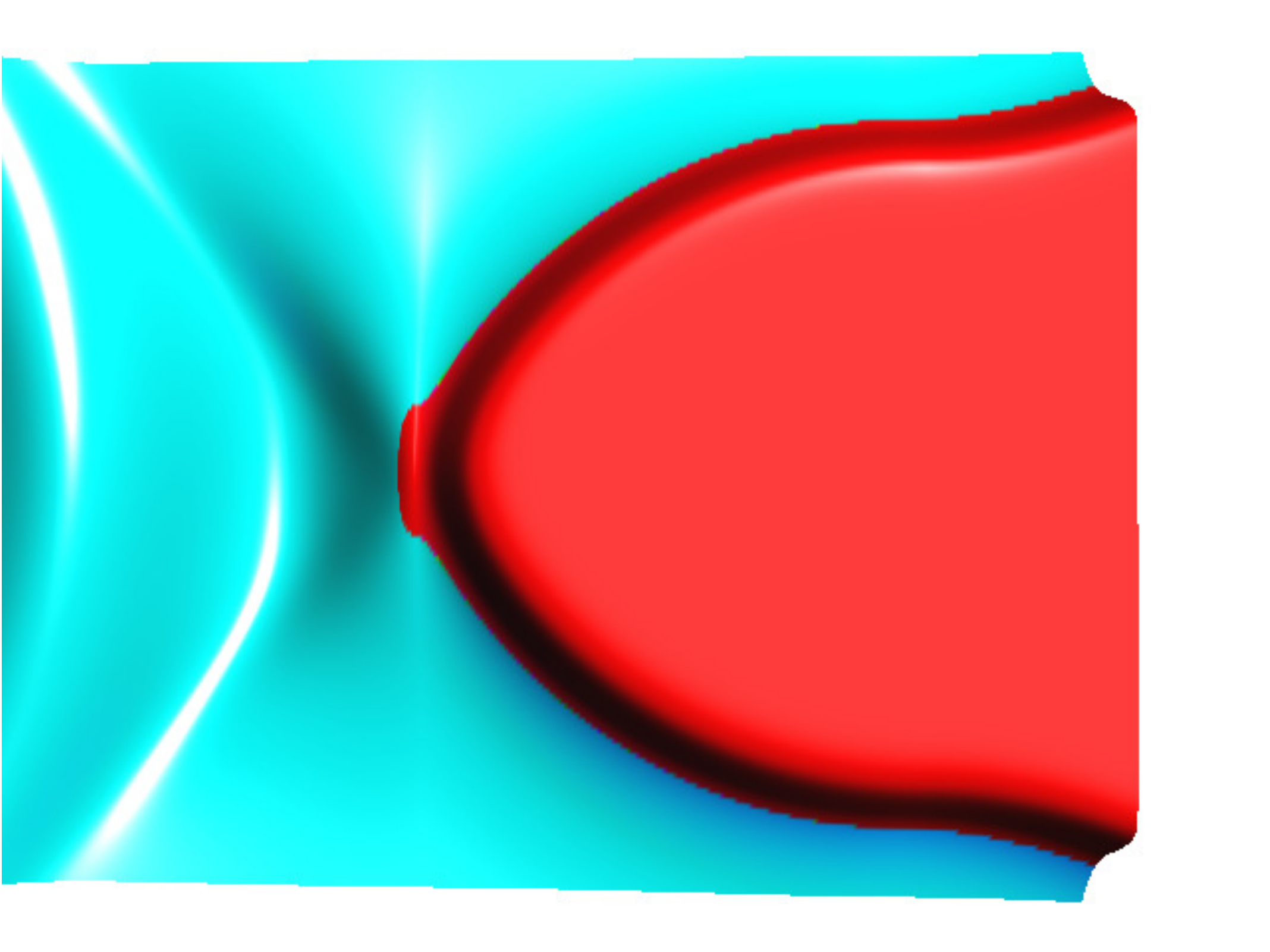}}
\put(15.7,142){\includegraphics[height=30.0\unitlength]{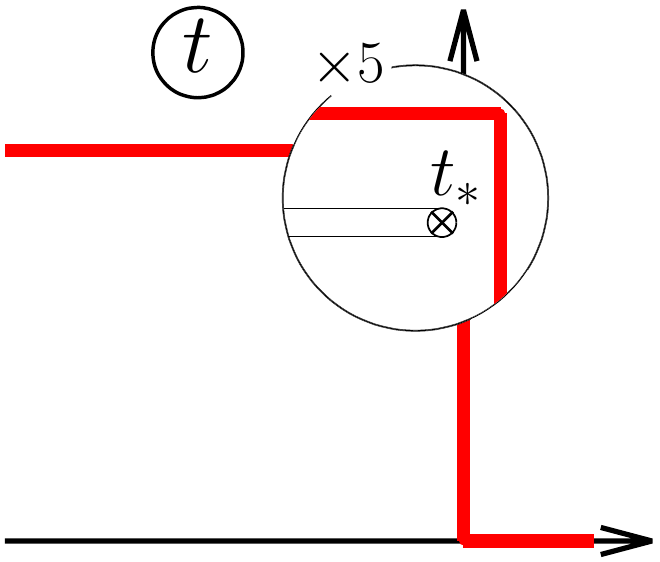}}
\thinlines
\thicklines
\put(40.8,140){\line(0,1){1}}
\put(23.68,140){\line(0,1){1}}
\put(23.68,141){\line(1,0){17.12}}
\linethickness{2mm}
\textcolor{white}{\qbezier(32.24,143)(32.24,148.9)(36.5,150)}
\thicklines
\qbezier(32.24,141)(32.24,148.9)(36.5,150)
\put(36.5,150){\vector(2,1){1}}
\put(-1,123){\begin{sideways}\large $x$\end{sideways}}
\put(17.5,101){$\mathrm{Re}\, t - \mathrm{Im}\, t$}
\put(2.7,134){4}
\put(2.7,124){0}
\put(2.1,114){-4}
\put(10.5,105.5){-12}
\put(20,105.5){-8}
\put(29,105.5){-4}
\put(38.8,105.5){0}
\put(8,30){\includegraphics[width=15\unitlength,angle=-90]{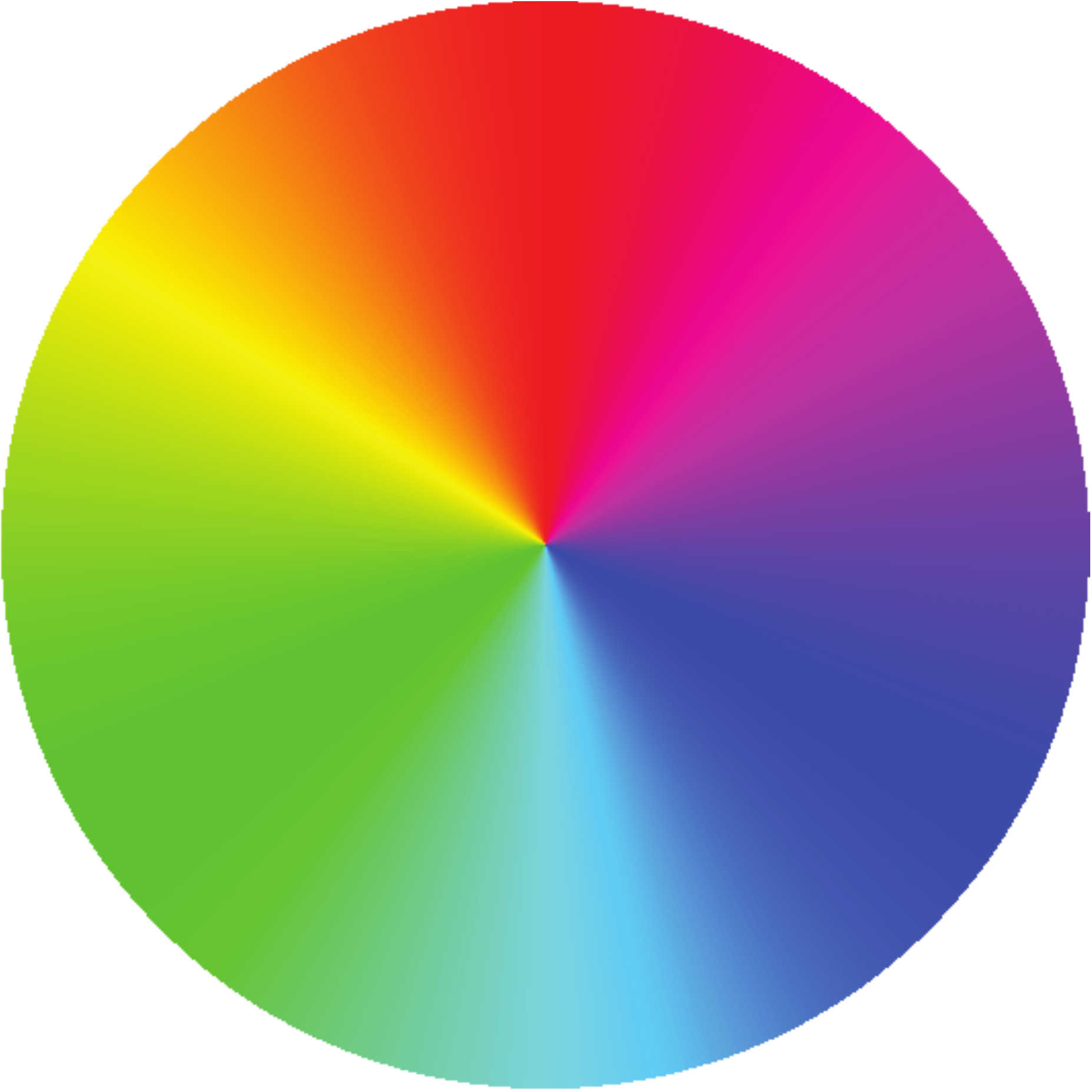}}
\put(3,9){\large $\mathrm{arg} (\phi_s-1)$}
\put(24.6,21.2){\Large $0$}
\put(3,21.7){\Large $\pi$}
\put(57,142){\includegraphics[width=35\unitlength,angle=-90]{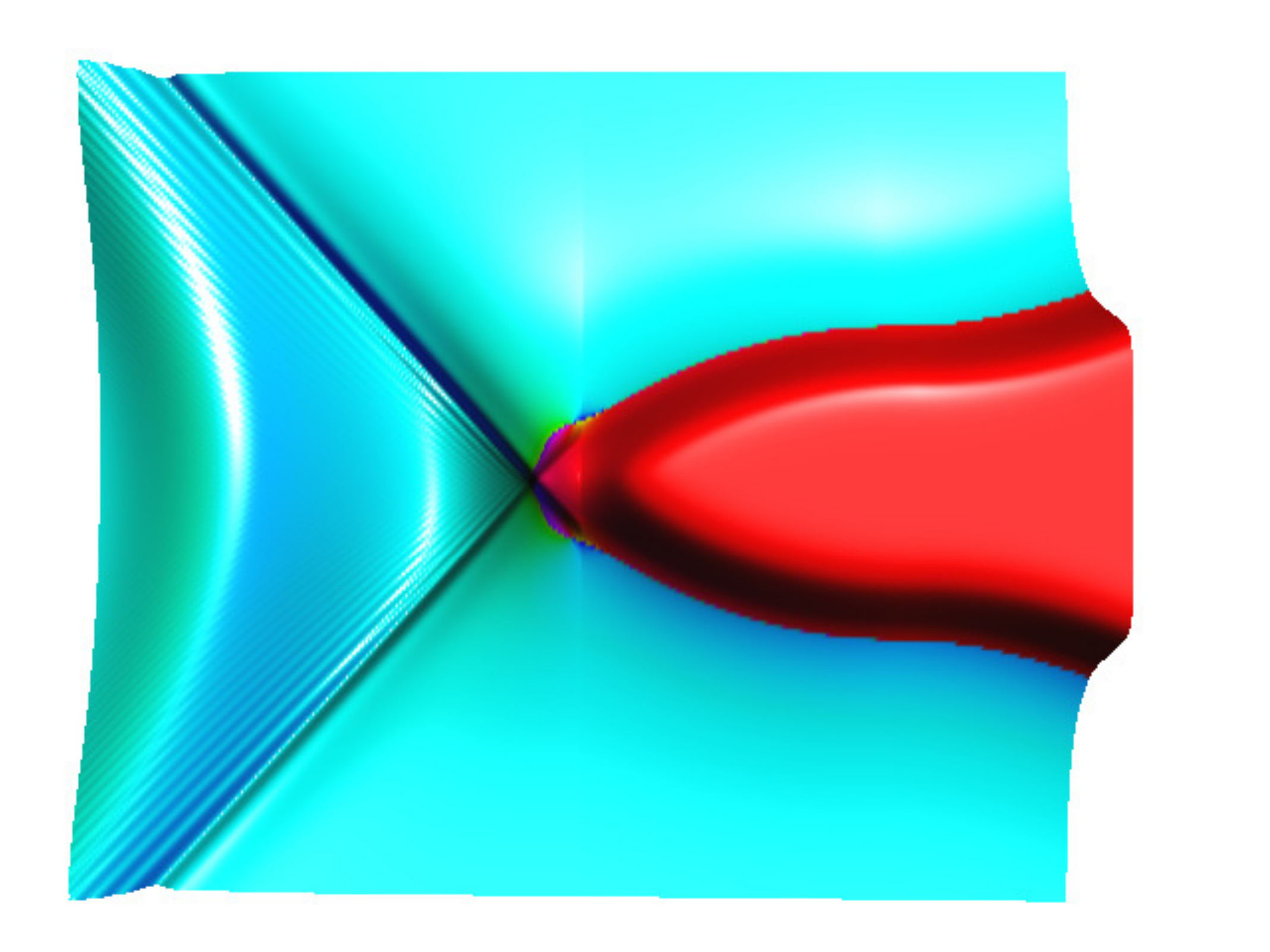}}
\put(69,142){\includegraphics[height=30\unitlength]{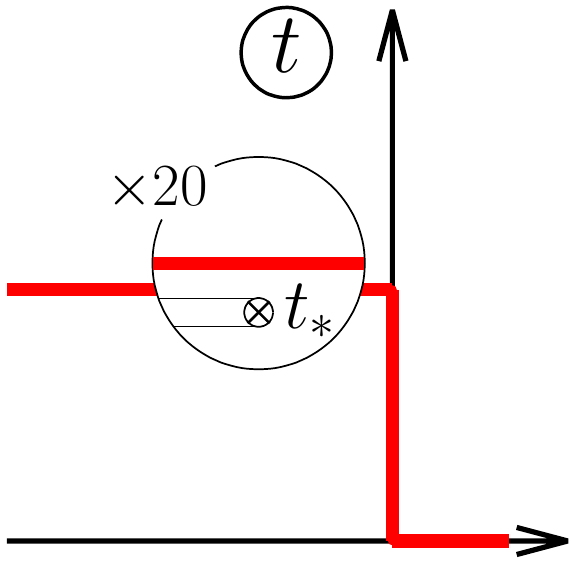}}
\linethickness{2mm}
\textcolor{white}{\qbezier(83.4,140.4)(83.4,146.4)(87,148)}
\thicklines
\qbezier(83.4,140.4)(83.4,146.4)(87,148)
\put(87,148){\vector(3,2){1}}
\put(77.5,139.4){\line(0,1){1}}
\put(89.3,139.4){\line(0,1){1}}
\put(77.5,140.4){\line(1,0){11.8}}
\put(55.5,134){4}
\put(55.5,124){0}
\put(54.5,114){-4}
\put(60.1,105.5){-12}
\put(69.9,105.5){-8}
\put(79.0,105.5){-4}
\put(88.6,105.5){0}
\put(51.0,123){\begin{sideways}\large $x$\end{sideways}}
\put(68,101){${\rm Re}\,t - {\rm Im}\,t$}
\put(106.8,145.65){\includegraphics[width=42\unitlength,angle=-90]{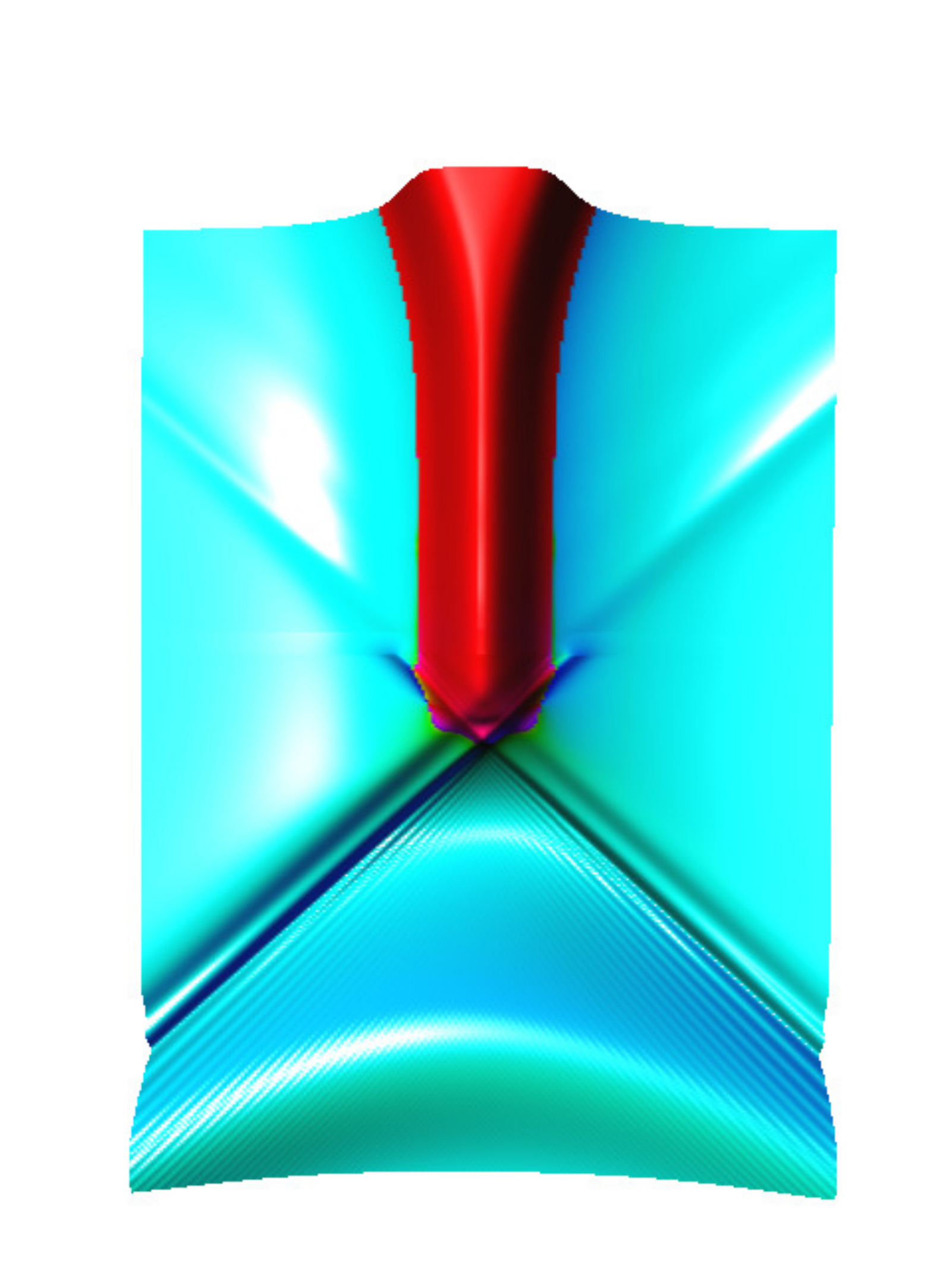}}
\put(109.6,142){\includegraphics[height=30\unitlength]{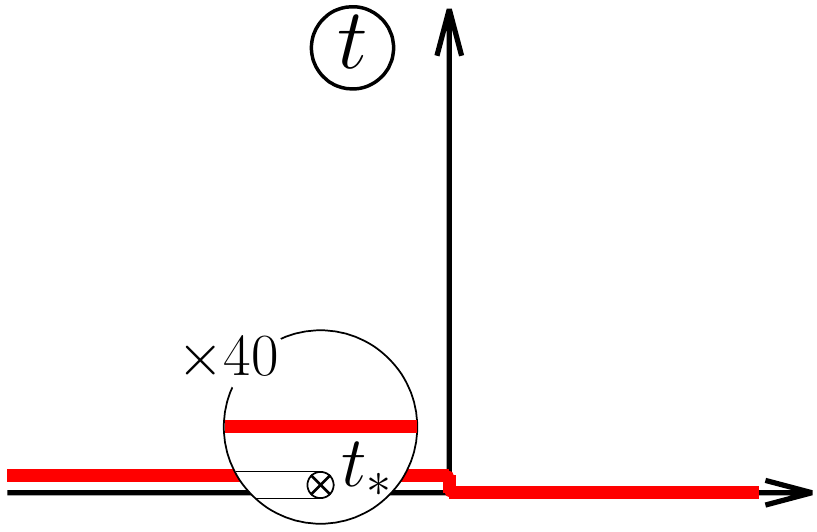}}
\put(123,101){${\rm Re}\,t - {\rm Im}\,t$}
\put(102.0,123){\begin{sideways}\large $x$\end{sideways}}
\put(106.7,135.5){4}
\put(106.7,123){0}
\put(106.1,111){-4}
\put(111.9,105.5){-8}
\put(123.2,105.5){-4}
\put(134.3,105.5){0}
\put(145.6,105.5){4}
\thicklines
\put(133.974,139.3){\line(0,1){1}}
\put(135,139.3){\line(0,1){1}}
\put(133.974,140.3){\line(1,0){1.026}}
\put(134.487,140.3){\vector(0,1){3}}
\put(106.3,49.2){\includegraphics[width=42\unitlength,angle=-90]{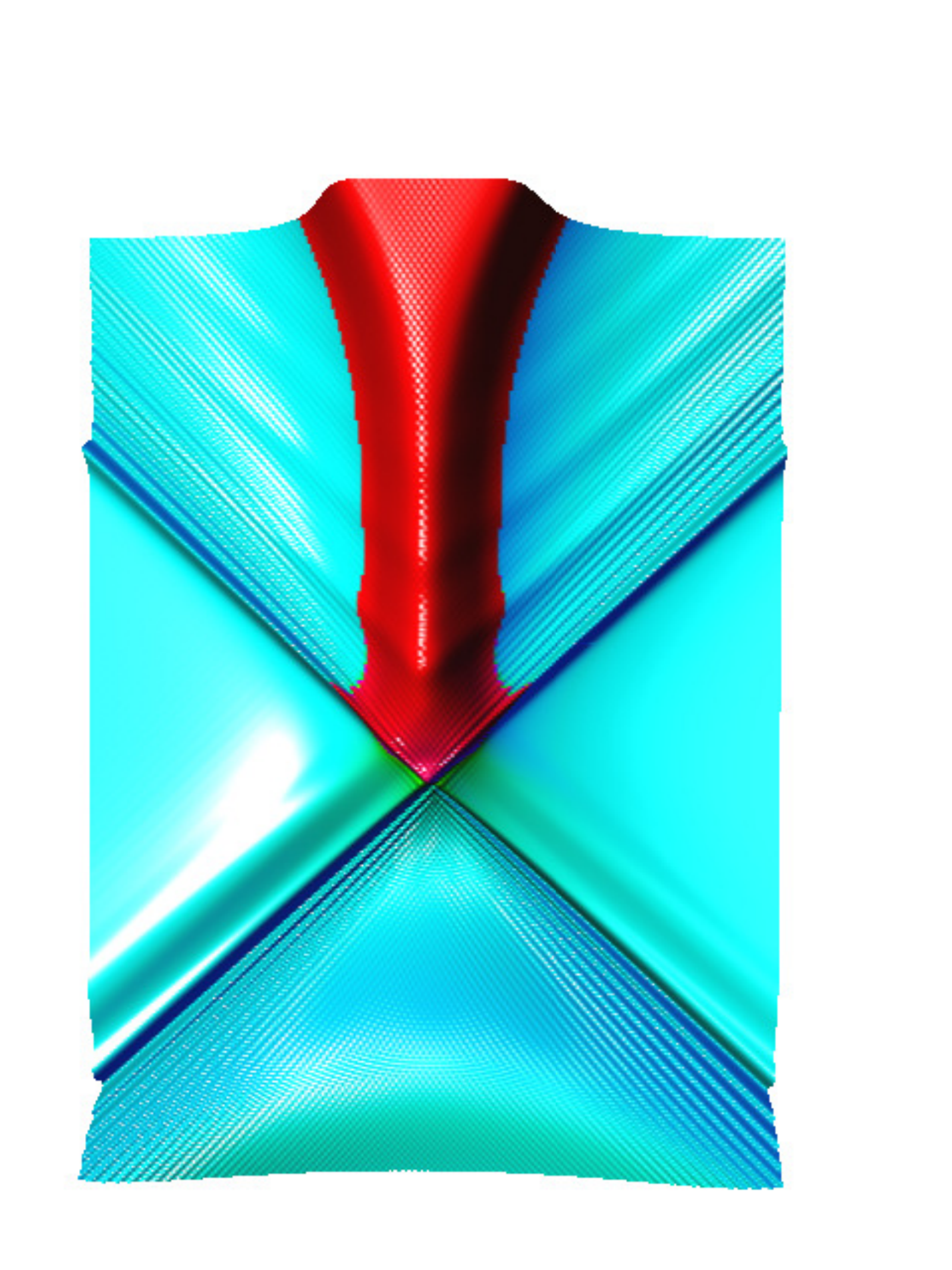}}
\put(109.6,48){\includegraphics[width=45\unitlength]{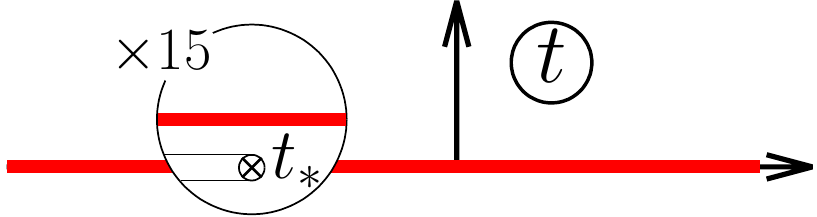}}
\thinlines
\put(134.2,10.7){0}
\put(123,10.7){-4}
\put(111.7,10.7){-8}
\put(145.3,10.7){4}
\put(106.7,41.5){4}
\put(106.7,29){0}
\put(106.1,16.8){-4}
\put(102,29){\begin{sideways}\large $x$\end{sideways}}
\put(123,5.5){${\rm Re}\,t - {\rm Im}\,t$}
\thinlines
\put(3,169){\Large (a)}
\put(53,169){\Large (b)}
\put(103,169){\Large (c)}
\put(103,58){\Large (d)}
\put(49,97.73){\line(0,1){78}}
\put(99,97.73){\line(0,1){78}}
\put(99,5){\line(0,1){60}}
\put(99,65){\line(1,0){41}}
\end{picture}}
\caption{{\it Central plot:} $(E,\, N)$ plane of initial
  data. Parameters of the numerical solutions are indicated by red
  (dark gray), purple and green (light gray) points. Circles with
  letters represent solutions from the insets $\mbox{(a)---(d)}$. The
  squared point corresponds to the real--time instanton with $T=0$ and
  $\theta=0.4$. 
  $\mbox{\it Insets~(a)---(d):}$ Three--dimensional plots of the numerical
  solutions $\mathrm{Re}\, \phi_s(t,x)$, where the abscisses $\mathrm{Re}\, t -
  \mathrm{Im}\, t$ parametrize the contours in the complex
  time plane. Color shows $\mathrm{arg}(\phi_s-1)$. Euclidean parts of
  the solutions are marked by the square brackets with arrows.  Above
  each 3D plot we draw the respective contour in the complex time
  plane, where the zoomed circular area with magnification factor
  displays the vicinity of the nearest singularity (crossed circle
  attached to the branch cut). The lattice size of all solutions is
  $N_t \times N_x = 3000 \times 150$.\label{fig:solutions}}  
\end{figure}

Our interest lies in the semiclassical results at high energies and
relatively small $N$. The well--known saddle--point configuration,
however, is the bounce $\phi_b(t,x)$ which satisfies the boundary value problem
\eqref{eq:54} at $T=+\infty$, $\theta=0$ and $E = N = 0$. We
continuously relate it to the high--energy solutions. Namely,
we find solutions at large $T$
by iterative numerical method using the bounce as the
zeroth--order approximation. After that we change $T$ and
$\theta$ in small steps and obtain one slightly deformed solution 
$\phi_s(t,x)$ of Eqs.~(\ref{eq:54}) at each step. We compute the energy $E$ and initial
particle number $N$  for every $\phi_s(t,\, x)$ by Eqs.~(\ref{eq:14}). An
exemplary set of numerical solutions is marked by points $(E,\, N)$ in
the central plot of Fig.~\ref{fig:solutions}. It was obtained by
decreasing $T$ at $\theta = 0$ (red/dark points), then increasing 
$\theta$ to $\theta=0.4$, and finally 
decreasing $T$ (green/light points) until the solution with $T=0$ and $\theta=0.4$ is
reached  (squared point). This procedure brings  us to the high--energy region of
interest. The other solutions in this region (not shown in
Fig.~\ref{fig:solutions}) are obtained in  similar manner, by
changing $(T,\, \theta)$ in small steps and numerically solving the
semiclassical equations~(\ref{eq:54}). 

Solutions $\phi_s = \phi_{rt}(t,\, x)$ with $T=0$ and arbitrary
$\theta$ are called ``real--time instantons.'' A representative
solution of this type corresponds to the squared point in 
Fig.~\ref{fig:solutions}. In the next Section we will find that 
the real--time instantons are radically different from the solutions
with $T\ne 0$. We will develop an adequate high--energy description of
collision--induced tunneling based on their properties.  

At low energies our semiclassical solutions reproduce the 
results of the previous Section. Indeed, Fig.~\ref{fig:theta}c compares
the semiclassical exponent (\ref{eq:15}) at $\theta=0$ (points) with the exponent in
Eq.~(\ref{eq:38}) (line), where the numerical values in 
Fig.~\ref{fig:theta}b are used. The two graphs coincide at $E \ll   
E_{cb}$ despite the fact that the perturbative results are obtained at $N=2$
whereas the semiclassical solutions involve large multiplicities $N
\sim 1/g^2$, see Fig.~\ref{fig:solutions}. This is because
$F_N(E)$ is independent\footnote{In the thin--wall approximation which
  works well~\cite{Demidov:2011dk} at $V(\phi_+) = -0.4$.}
of $N$ at $E < E_{cb}$~\cite{Rubakov:1992gi}, so we do not have to
continue it to $g^2 N \to 0$. Importantly, the perturbative graph in
Fig.~\ref{fig:theta}c becomes negative at $E \approx 0.8E_{cb}$ indicating
apparent violation of unitarity. At these energies the perturbative
expansion of Sec.~\ref{sec:grow-ampl-backgr} is not reliable unlike
the semiclassical method of this Section. 

Our numerical results show a dramatic change in the form of the
saddle--point solutions at $E\approx E_{cb}$. The low--energy
solutions in Figs.~\ref{fig:solutions}a,b resemble the bounce: they
contain long Euclidean parts and describe creation of true vacuum bubbles 
(the latter are shown by red (dark) in the figures). Waves in the left parts of the
plots represent initial particles. In contrast, at $E>E_{cb}$ the
initial waves are sharper and the bubbles are 
smaller, see Figs.~\ref{fig:solutions}c,d. Another property of
high--energy solutions is small durations of their Euclidean
evolutions and, nevertheless, nonzero values of the suppression
exponents\footnote{Exponentially suppressed transitions at $E>E_{cb}$
  are called ``dynamical tunneling''~\cite{Miller,Heller} to
  distinguish from the potential tunneling at low
  energies.}~(\ref{eq:15}) thanks to complex--valued $\phi_s(t,x)$.
In particular, the real--time instanton at $T=0$ evolves entirely 
along $t\in \mathbb{R}$. We will argue that this feature guarantees
stable perturbative expansion at high energies.

One wonders how the durations of Euclidean evolutions can be of any
physical meaning: they are not even explicitly specified in
Fig.~\ref{fig:theta}a. Indeed, one expects that $\phi_s(t,x)$ are
analytic functions of time and can be continued to any
contours. But in fact, the solutions and, in particular,
the bounce have branch--cut 
singularities starting at $t = t_*$ which  separate their time contours from the
real time axis, see Fig.~\ref{fig:theta}a. We compute positions of these
singularities for the solutions (a)---(d) in Fig.~\ref{fig:solutions}. To
this end we continue $\phi_s(t,\, x)$ to complex $t$ and find the
points 
$t=t_*$ where $|V(\phi_s)| = \infty$, see the zoomed areas in the
respective $t$--planes. The singularities prevent us from continuing
the low--energy solutions to the real time axis. We find, however, that
$\mathrm{Im}\, t_*$ decreases with energy and becomes
negative for the solution (d) in Fig.~\ref{fig:solutions}.  The latter
solution can be considered in real time, as well as the real--time
instanton at $T=0$.

We find the semiclassical solutions at $g^2 N = 2$ (not shown in
Fig.~\ref{fig:solutions}) and plot the respective 
suppression exponent $F_N(E)$ in Fig.~\ref{fig:potential_energy}c
(thick line in the left part of the lower panel). The latter exponent
monotonously decreases with energy 
until the point $T=0$ and $E = E_{rt}(N)$ is reached\footnote{Data
  lines are not shown at high energies where the
  numerical errors are   large. To decrease the computational cost, we
  find the solution at $E = 
  E_{rt}$ at high precision (next Section) and fill the gap with thin
  interpolating line.}.
The region of few--particle initial states $N \ll 1/g^2$ cannot be
directly addressed at the present--day computers. So, we extrapolate
numerical results into this region. Since $\mathrm{e}^{-\theta}$
analytically enters the semiclassical equations \eqref{eq:54}, the
particle number $N$, action $S[\phi_s]$, 
and the 
last term in Eq.~(\ref{eq:15}) have regular Taylor expansions in this
parameter. Moreover, $\mathrm{e}^{-\theta} \to 0$ leads to Feynman
boundary conditions at $t\to -\infty$ and therefore to $g^2 N \to
0$, see Eq.~(\ref{eq:39}). Substituting all Taylor expansions into 
Eq.~(\ref{eq:15}) and using Eq.~(\ref{eq:16}), one finds that 
\[F_N(E) + g^2\theta N + g^2 N = F(E) + O(g^4N^2)\;.\] We fit numerical data
for the left--hand side of this equality with function $F(E) + d(E)
\cdot g^4 N^2$ and obtain the suppression exponent $F(E)$ 
(solid line in the left upper part of
Fig.~\ref{fig:potential_energy}c).  Numerical error of this 
procedure is expected to be smaller than $5\%$. Our results show that
the two--particle exponent $F(E)$ is also a decreasing function 
of energy. 

We finish this Section by remarking that the above semiclassical
solutions describe classically forbidden transitions from the initial
states with relatively small multiplicities $N$. Increasing $N$, one 
reaches the states decaying classically (region ``$F_N=0$'' in
Fig.~\ref{fig:solutions}).  Classical transitions in the
model~(\ref{eq:9}) were studied in Ref.~\cite{Demidov:2011eu}, see 
Refs.~\cite{Dutta:2008jt, Romanczukiewicz:2010eg, Lamm:2013ye} for the 
related work.  

\section{Real--time instantons}
\label{sec:real-time-instatons}
\begin{figure}[t]
\vspace{3mm}

\hspace{3.5cm}~~~ \hspace{8.5cm}(b)

\vspace{-0.7cm}

\unitlength 1mm
\centerline{\begin{picture}(50,55)(0,-9)
\put(-0.2,50.2){\includegraphics[width=50\unitlength,angle=-90]{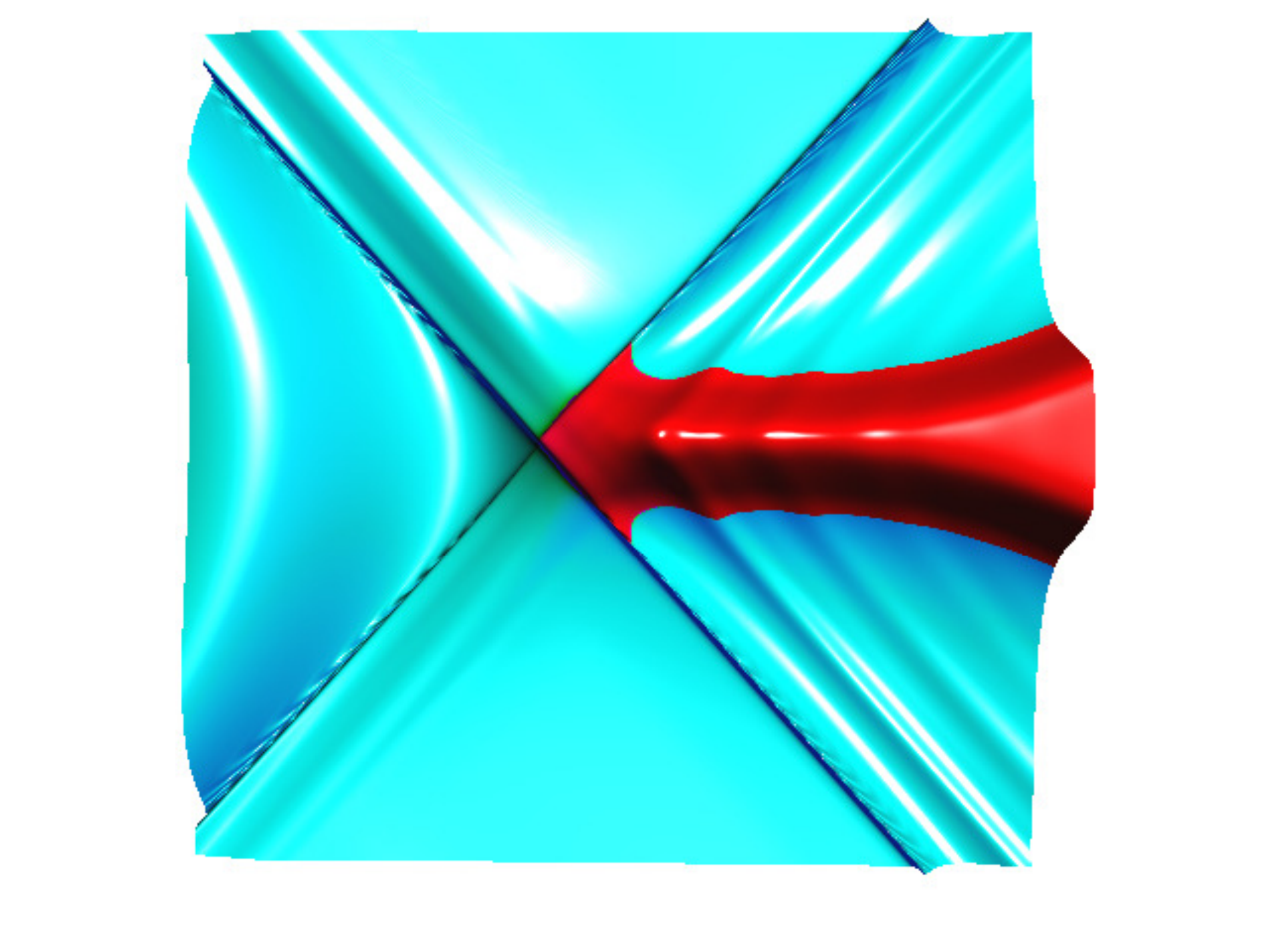}}
\put(30,52){(a)}
\put(11.3,0.5){-8}
\put(24.3,0.5){-4}
\put(37.35,0.5){0}
\put(31,-6){\Large $t$}
\put(50.4,0.5){4}
\put(4.5,11.1){-4}
\put(6,25.9){0}
\put(-0.5,25.7){\Large \begin{sideways}$x$\end{sideways}}
\put(6,40.9){4}
\end{picture}\hspace{2cm}
\includegraphics[width=7.5cm]{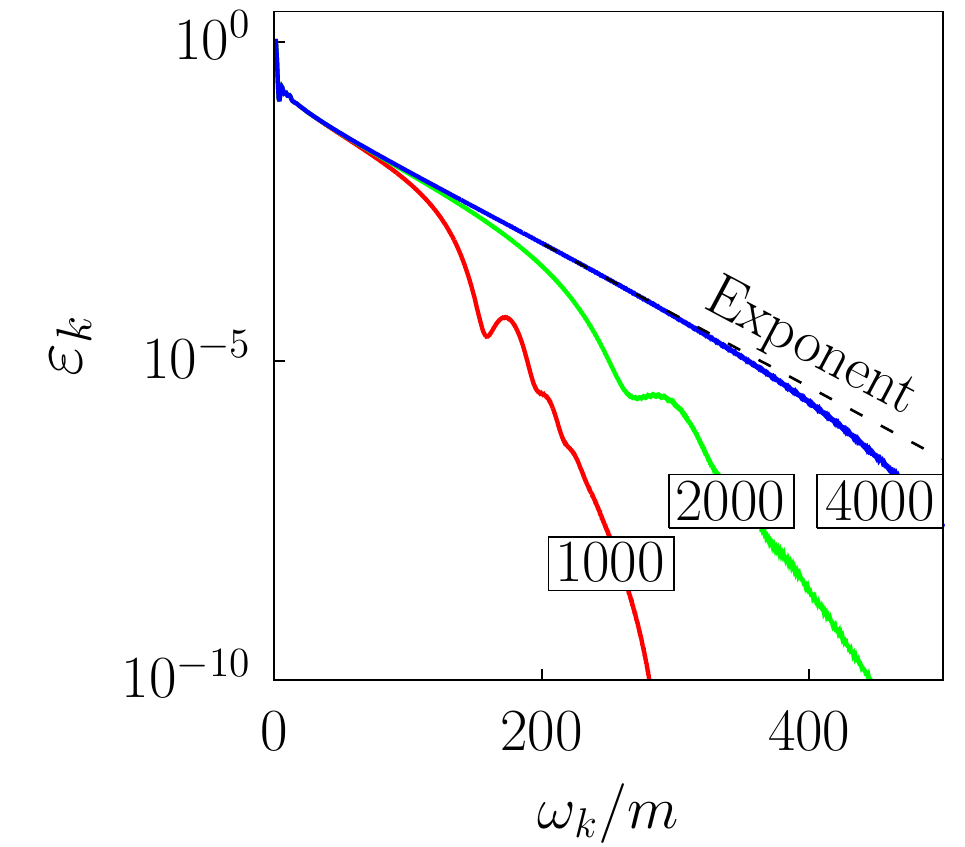}\hspace{3mm}}

\vspace{-2mm}

\caption{(a) Real--time instanton $\phi_{rt}(t,x)$ and
  (b) its initial energy distributions $\varepsilon_k$ at different
  $N_x$ (numbers in boxes). In both figures $\theta = 0.199$,
  $(E,\, N) \approx 
  (2.4\, E_{cb},\, 3.5/g^2)$, $N_t = 11000$. Figure (a) uses colors of
  Fig.~\ref{fig:solutions} and $N_x=2000$.\label{fig:rt}} 
\end{figure}
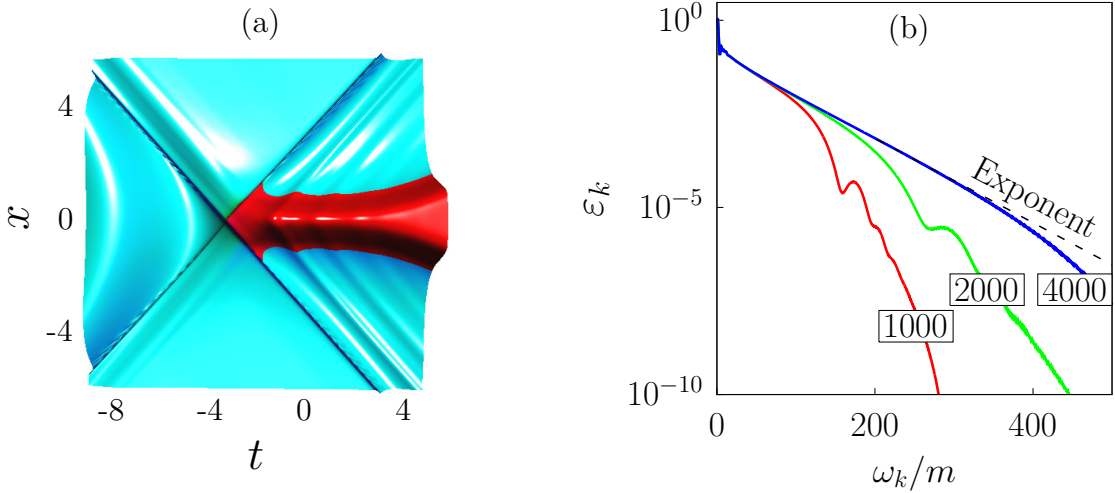

The real--time instantons, i.e.\ solutions of the saddle--point
equations at $T=0$ and arbitrary $\theta$, are special in
many respects (see one of them in Fig.~\ref{fig:rt}a). At a given
$N$ they have the highest energies $E = E_{rt}(N)$ 
and smallest suppression exponents $F_{min}(N) \equiv
F_{N}(E_{rt}(N))$ considered so far. Indeed, all solutions at smaller
energies have positive parameter $T$: we obtained them by lowering
$T$ from $T = +\infty$ to $T=0$. Since $\partial_E F_N = -2g^2T$, the
exponent $F_N(E)$ decreases with energy and reaches a local minimum at
$E = E_{rt}(N)$. In Sec.~\ref{sec:transitions-at-ee_rt} we will argue
that this minimum is global by showing that the suppression exponent
is energy--independent at $E>E_{rt}(N)$. 

The above property implies that each real--time instanton describes
collision--induced transitions from the states with fixed multiplicity
$N$ and arbitrary energies. Indeed, the probability 
\[
\sigma^{max}_{N}  = \int_0^\infty dE\; \mathrm{e}^{-F_N(E)/g^2} \;,
\]
receives dominant contribution near the saddle point $E =
E_{rt}(N)$ corresponding to the real--time instanton. One finds
$\sigma^{max}_N \approx \mathrm{e}^{-F_{min}(N) / g^2}$, where
$F_{min}(N)$ is computed using $\phi_{rt}(t,\, x)$. 

In the previous Section we obtained the real--time instantons from the
larger family of semiclassical solutions with arbitrary $T$ and
$\theta$. One can compute $\phi_{rt}$ directly by solving the
semiclassical equations (\ref{eq:56}), (\ref{eq:57}) with the initial
condition 
\begin{equation}
\label{eq:55}
a_k = \mathrm{e}^{-\theta} b_k\;, 
\end{equation}
cf. Eq.~(\ref{eq:39}). Then the parameter $N$, energy $E_{rt}(N)$ and
minimal suppression exponent $F_{min}(N)$ are given by the standard
expressions (\ref{eq:14}),~(\ref{eq:15}).

A convenient and important feature of the real--time instantons came
as a surprising numerical fact in Fig.~\ref{fig:solutions}: they
are defined  
along the real 
time axis $t\in \mathbb{R}$. Technically, this property is related to
the initial condition (\ref{eq:39}) and its consequence (\ref{eq:55}). Indeed, the positive--
and negative--frequency terms in the integrand of Eq.~(\ref{eq:17})
are of order $\tilde{b}_k^* \equiv b_k^* \,\mathrm{e}^{-\omega_k
  T_{AB}}$ and 
$\tilde{a}_k \equiv a_k\, \mathrm{e}^{\omega_k T_{AB}}$, where $T_{AB}
= \mathrm{Im}\, t_{AB}$ is the height of the time
contour. The integrals of these terms in
Eq.~(\ref{eq:17}) converge faster and slower at higher
$T_{AB}$, respectively. At $T_{AB} = T$ the terms are of the same
order because $\tilde{a}_k = \mathrm{e}^{-\theta} \, \tilde{b}_k$ due to the
initial condition (\ref{eq:39}). This ``optimal''
contour lies right in the middle between the singularities of the 
solution: at somewhat higher or lower $T_{AB}$ one  of the integrals
in Eq.~(\ref{eq:17}) diverges signaling that the contour hits the
singularity. We use $T_{AB}=T$ in numerical calculations and show the
corresponding contour
in Figs.~\ref{fig:solutions}a--d. At $T\to 0$ the optimal contour coincides
with the real time axis.

The above argument turns quantitative if we use the high--frequency
asymptotics of the solution. Namely, consider the energy of modes with 
the wave number $k$ in Eq.~(\ref{eq:14}), 
\begin{equation}
\label{eq:40}
 \varepsilon_k \equiv \frac{a_k b_k^*}{4\pi}\;, \qquad\qquad g^2 E =
 \int dk \, \varepsilon_k\;.
\end{equation}
In Appendix~\ref{sec:semicl-solut-at} we demonstrate that any smooth
solution has exponential asymptotics,
\begin{equation}
\label{eq:18}
\varepsilon_k \to \varepsilon_0 \,
\mathrm{e}^{-2\omega_k T_*} \qquad\qquad \mbox{as}\qquad k \to +\infty\;,
\end{equation}
where $T_*$ is a parameter of the solution. Extracting $a_k$ and $b_k^*$ 
from Eq.~(\ref{eq:18}) and  initial condition (\ref{eq:39}), one finds that the integral in
Eq.~(\ref{eq:17}) converges at $|T_{AB} - T|< T_*$ and diverges
otherwise.  Thus, $T_*$ is the distance from the optimal 
contour to the nearest singularity of the solution, $T_* = |T-\mathrm{Im}\, t_*|$. 

In Fig.~\ref{fig:rt}b we plot energy distributions $\varepsilon_k$ 
for the real--time instanton with $\theta=0.199$ at different
$N_x$. At larger $N_x$ and $\omega_k \gg m$ the graphs approach
Eq.~(\ref{eq:18}) with $T_* 
\approx 0.013$. This indicates that the continuum limit of our
numerical solution is a smooth configuration with singularities
at finite distances $T_*$ from the real time axis.

\begin{figure}[t]

\vspace{1.2cm}
\hspace{4.3cm} (a) \hspace{6.7cm} (b)

\vspace{-1.2cm}
\begin{minipage}{8cm}
\hspace{2mm}\includegraphics[height=6.7cm]{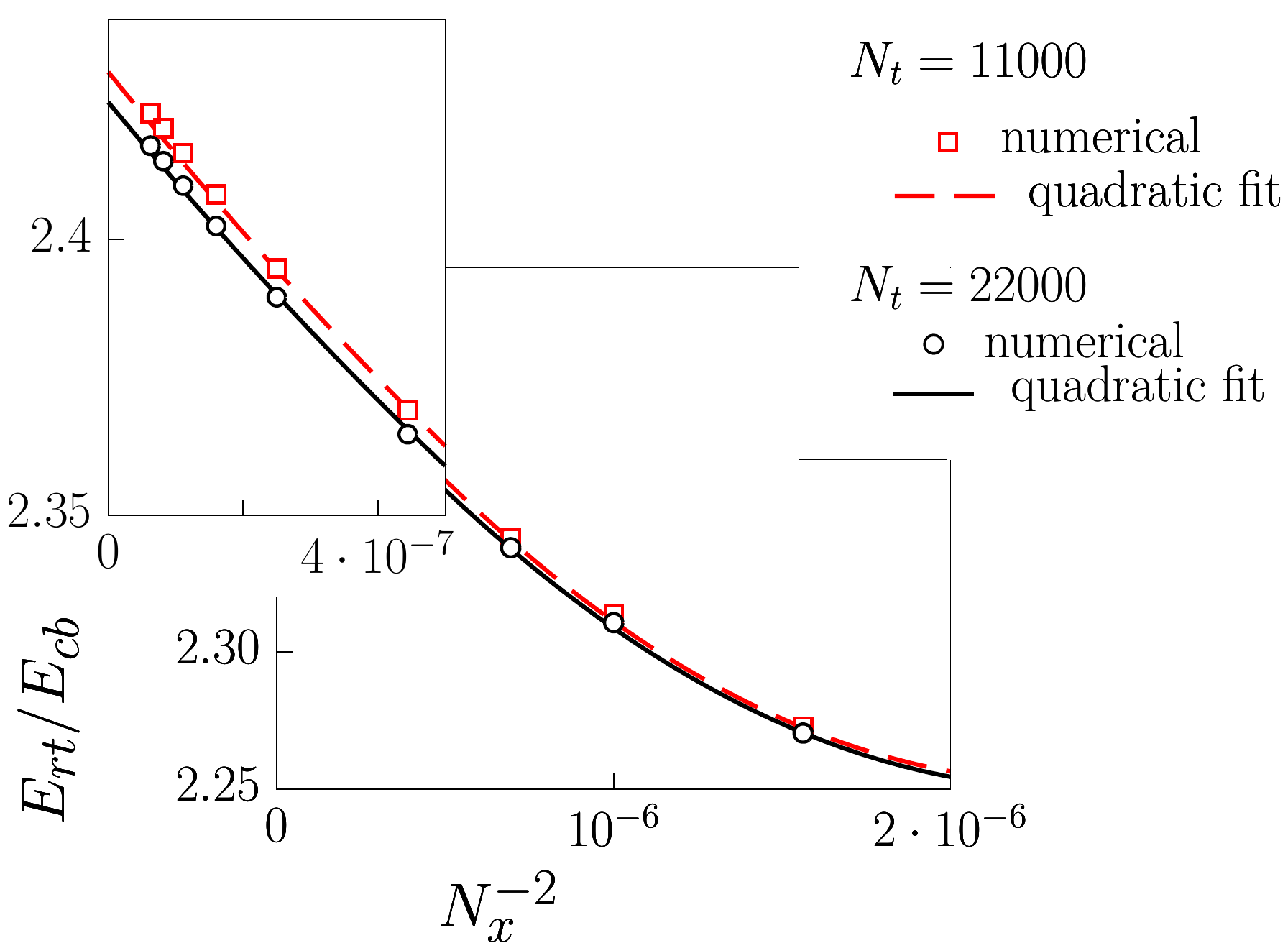}
\hspace{2mm}
\end{minipage} \hspace{15mm}
\begin{minipage}{9cm}
\includegraphics[height=6.7cm]{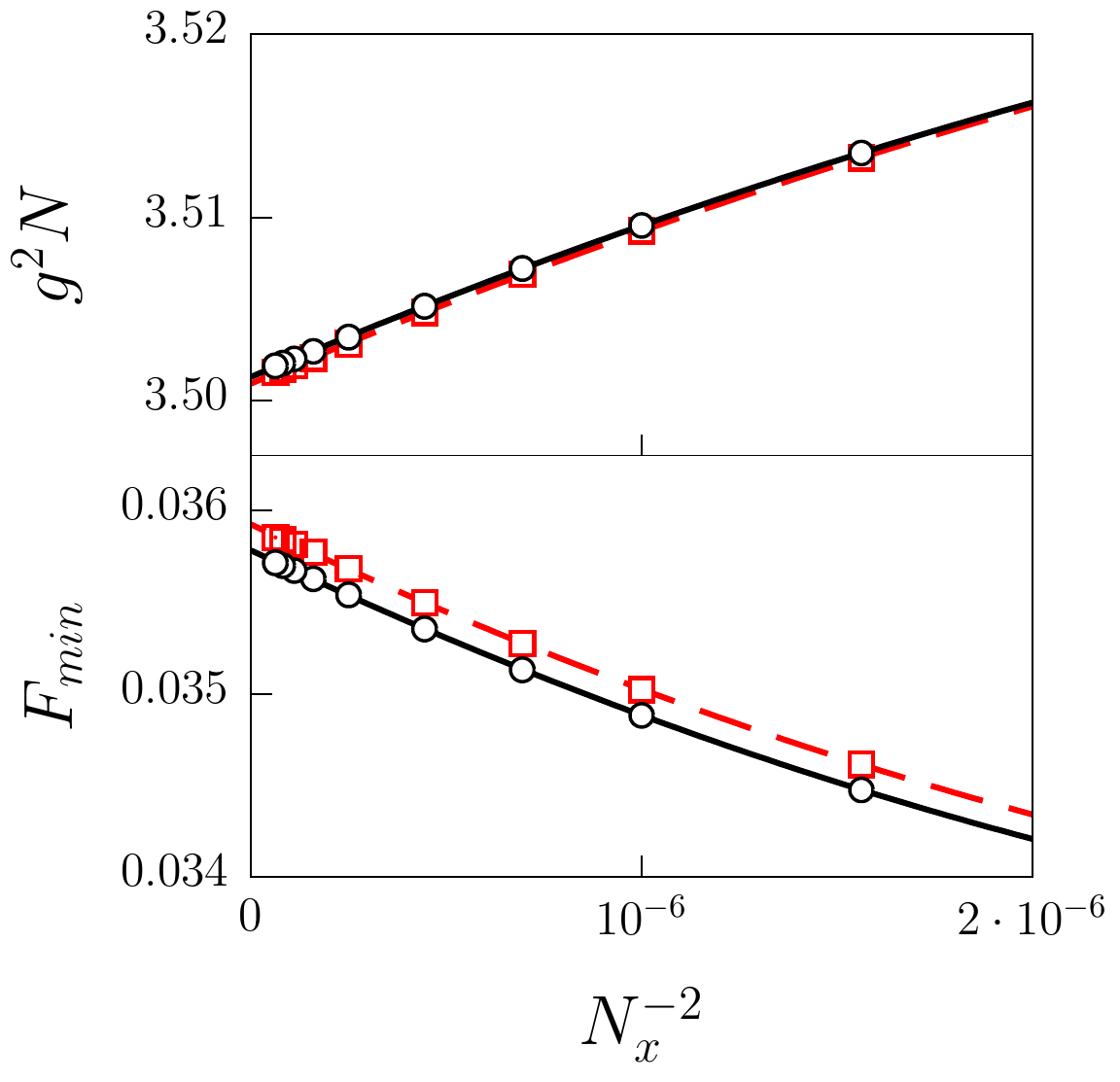}
\end{minipage}

\vspace{-2mm}
\caption{Continuum limits $N_x \to +\infty$ of the (a) real--time
  instanton energy $E_{rt}$, (b) its  initial particle number $N$ and
  suppression exponent $F_{min}$; $\theta=0.199$. We fit the data points
  with quadratic functions of $\delta \equiv N_x^{-2}$ (lines) in the
  range $N_x = 800 \div 4000$.\label{fig:continuum}}
\end{figure}
Note that the latter fact is important because our conclusions about
existence and properties of the real--time instantons rely on numerical
calculations. In Fig.~\ref{fig:continuum} we study the continuum limit 
in more detail. If $\phi_{rt}(t,\, x)$ are smooth configurations, the
numerical errors are expected to be polynomials in $\delta \equiv
N_x^{-2} \propto \left(\Delta x\right)^2$ and $N_t^{-2}$ because we use
the second--order finite--difference methods. This is
indeed the case: lattice values of $E$, $N$ and $F_{min}$ (points  
in Fig.~\ref{fig:continuum}) are not sensitive to $N_t$ and well
approximated by quadratic functions of 
$\delta$ (lines). At large $N_x$ all numerical errors are proportional
to $\delta$, see the inset in Fig.~\ref{fig:continuum}a. Once again we
conclude that our lattice solutions with $T=0$ have well--defined
continuum limits.

\begin{figure}[t]
\vspace{5mm}

\hspace{4.2cm} (a) \hspace{7.8cm} (b)

\vspace{-7mm}
\centerline{\includegraphics[width=7.5cm]{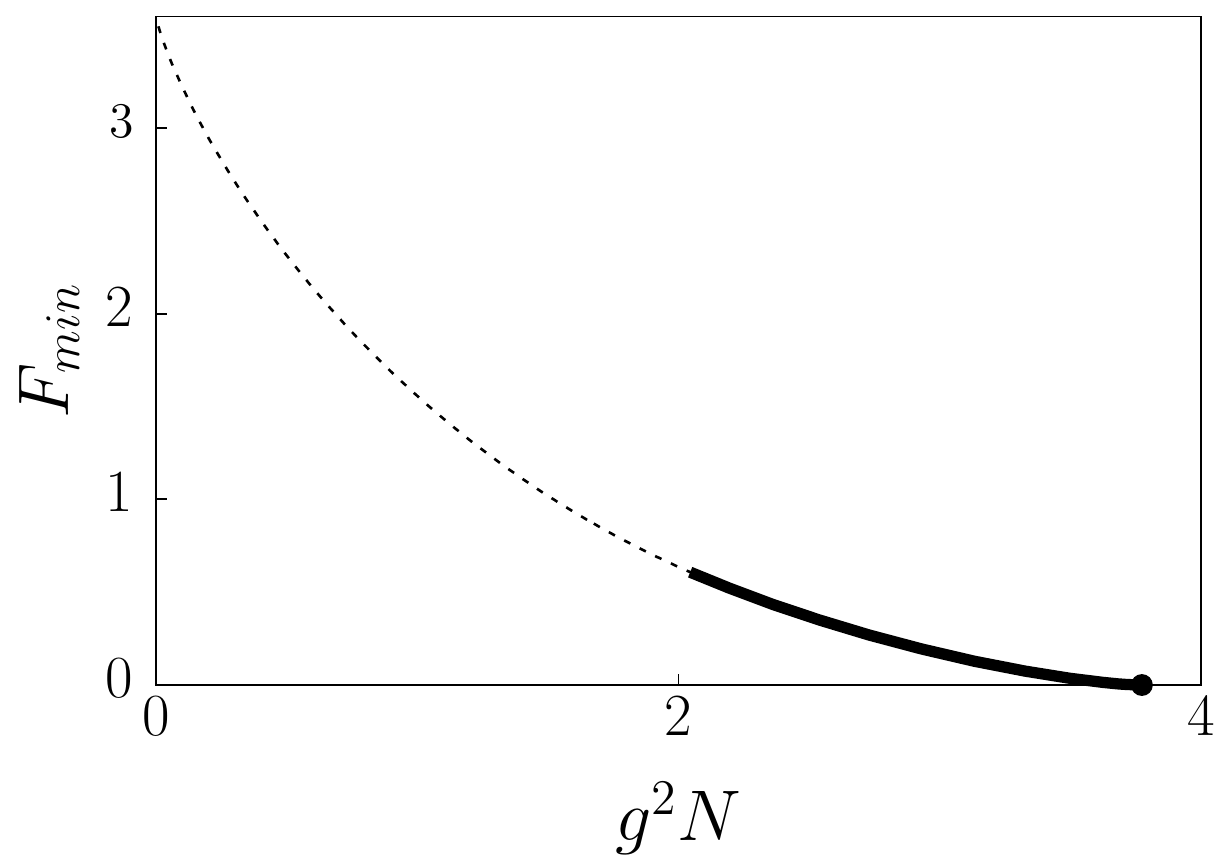}\hspace{1cm}
\includegraphics[width=7.5cm]{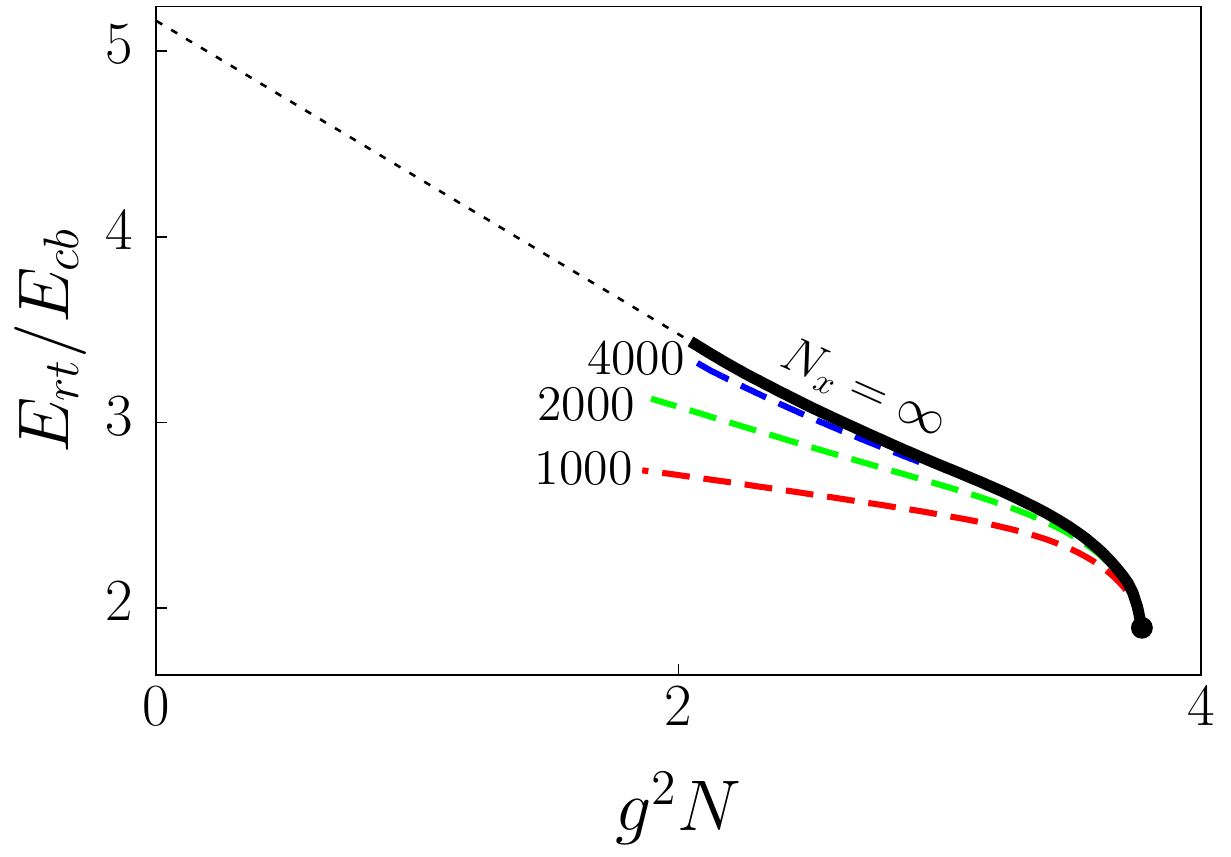}}
\caption{(a) Suppression exponent $F_{min}(N)$ and (b) energy
  $E_{rt}(N)$ of the real--time instantons. Dashed lines in Fig.\ (b)
  are the lattice results at $N_t = 11000$ and different 
  $N_x$ (numbers near the lines). Thick solid lines are obtained by
  extrapolating results to $N_x \to +\infty$.\label{fig:Ert}}
\end{figure}
Numerical results for the minimal suppressions $F_{min}(N)$ and
respective energies $E_{rt}(N)$ are shown in
Fig.~\ref{fig:Ert}. Dashed lines in Fig.~\ref{fig:Ert}b  
are the lattice results\footnote{In Fig.~\ref{fig:Ert}a they are
  indistinguishable from the continuum limit.}, while 
the solid lines represent continuum limits obtained by
quadratic extrapolations in $\delta = N_x^{-2}$ to $\delta = 0$,
cf. Fig.~\ref{fig:continuum}. Points $E=E_{rt}$ in
Fig.~\ref{fig:potential_energy}c are extracted from
Fig.~\ref{fig:Ert}. In particular, we obtain results for the few--particle
initial states $g^2 N \ll 1$ by extrapolating $E_{rt}(N)$ to $g^2 N=0$ 
with linear function, and $F_{rt}(N)$ by the method described in the
end of Sec.~\ref{sec:from-euclidean-real}, see the dotted lines in
Fig.~\ref{fig:Ert}. The accuracy of our result for $F_{rt}(0)$ is
better than $5\%$, while extrapolation of energy should be 
considered as illustrative. In particular, we cannot completely exclude the
possibility that $E_{rt} \to +\infty$ at $g^2 N \to 0$. Nevertheless,
it is likely that the point $E = E_{rt}$ exists for the few--particle
initial states, since it does for the multiparticle ones,
cf. Refs.~\cite{Levkov:2004tf, Levkov:2004ij}.

\section{Transitions at $E>E_{rt}$}
\label{sec:transitions-at-ee_rt}
Numerical solutions at $E > E_{rt}(N)$ look similar to the real--time
instantons, cf. Figs.~\ref{fig:Tm0}a and~\ref{fig:rt}a. But in fact
they are fundamentally different: we are going to demonstrate that
they do not have continuum limits and lead to
energy--independent suppression exponent $F_{N}(E) = F_{min}(N)$.

\begin{figure}[t!]

\hspace{8.5cm}  (b) \hspace{4.9cm} (c)
\vspace{-3mm}

\unitlength=1mm
\begin{picture}(50,55)(0,-4.5)
\put(1.75,48.4){\includegraphics[width=47.5\unitlength,angle=-90]{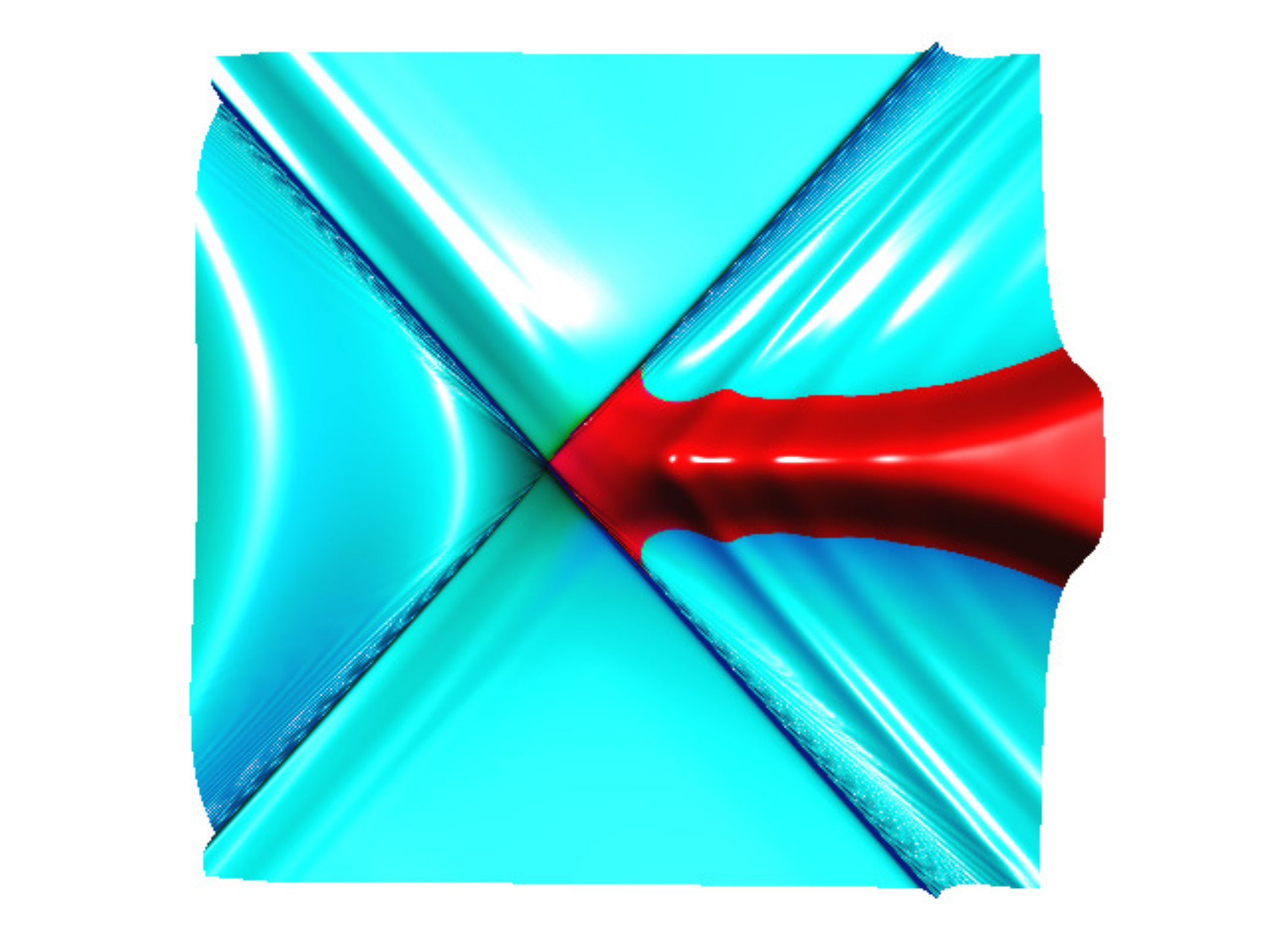}}
\put(30,50){(a)}
\put(12.8,0.5){-8}
\put(25.2,0.5){-4}
\put(37.6,0.5){0}
\put(31,-5){\Large $t$}
\put(50.3,0.5){4}
\put(6.6,10){-4}
\put(8,24.1){0}
\put(2,24){\Large \begin{sideways}$x$\end{sideways}}
\put(8,38){4}
\end{picture}
\hspace{0.8cm}
\includegraphics[width=4.89cm]{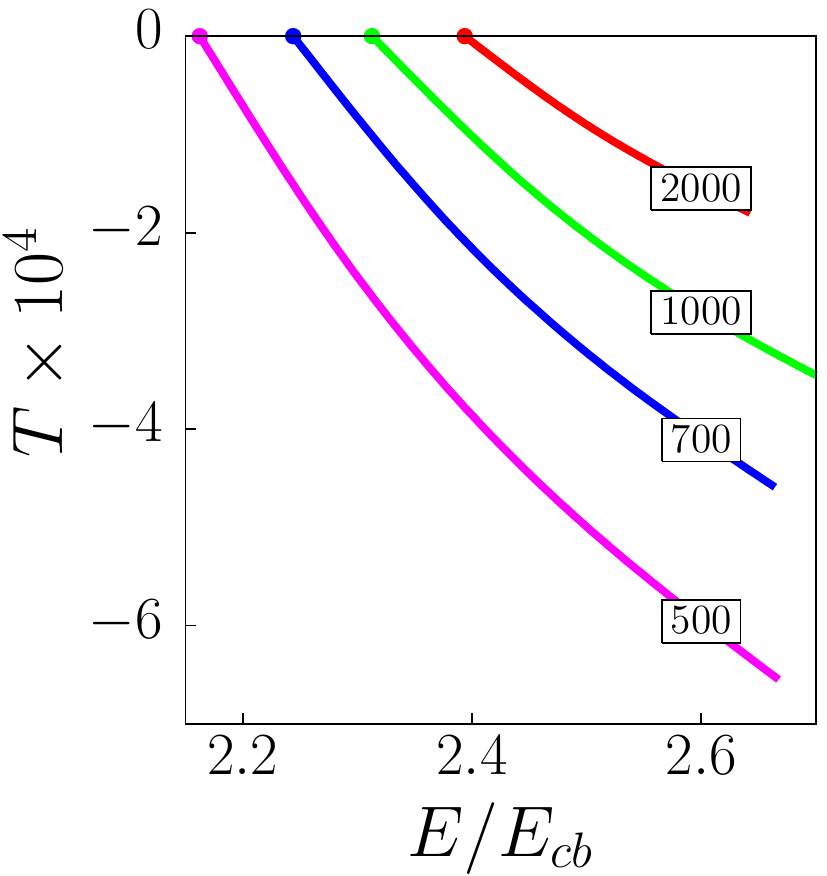}\hspace{5mm}
\includegraphics[width=5.11cm]{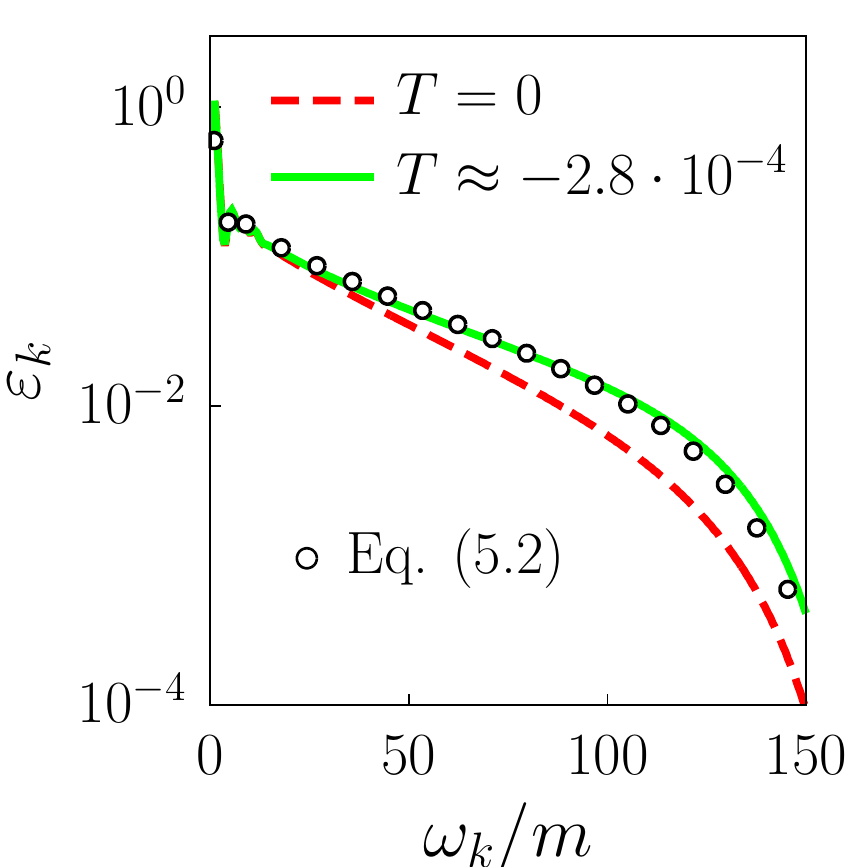}

\caption{(a) Numerical solution $\phi_s(t,x)$ at $E \approx 2.6 E_{cb} >
  E_{rt}(N)$. The other parameters are $g^2 N \approx 3.5$, $T \approx -2.8\cdot
  10^{-4}$,  and $N_x = 1000$. (b) Parameter $T$ in the region $E > E_{rt}(N)$  for
  different $N_{x}$ (numbers in boxes). (c) Energy
  distributions for the solution (a) and the real--time
  instanton with $N_x = 1000$ from Fig.~\ref{fig:rt}b. In all figures 
  $\theta = 0.199$ and $N_t = 11000$.\label{fig:Tm0}}
\end{figure}
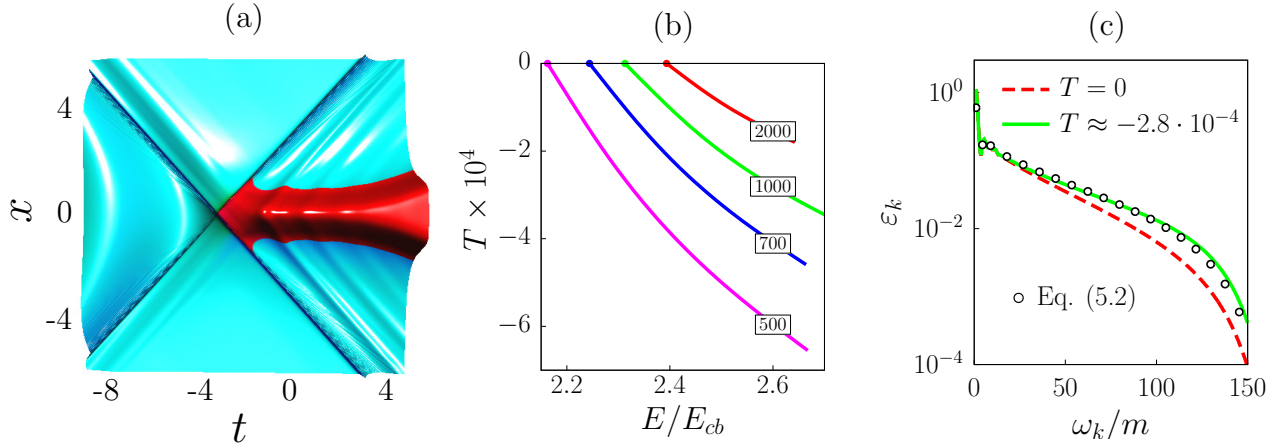
\begin{wrapfigure}[9]{r}{5.1cm}
\hspace{-1mm}\includegraphics[width=4.9cm]{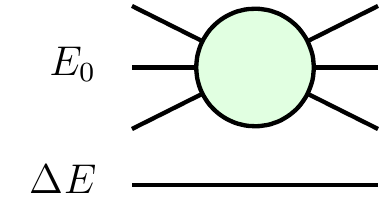}

\vspace{-2mm}
\caption{Exclusive transition at $E>E_{rt}(N)$.\label{fig:diagram}}
\end{wrapfigure}
To begin with, we find that the lattice values of the parameter $T$
monotonously decrease with energy and become negative at $E >
E_{rt}(N)$, see Fig.~\ref{fig:Tm0}b. Then Eq.~(\ref{eq:16}) implies that 
the exponent $F_N(E)$ reaches minimum at $E=E_{rt}(N)$ and increases
at 
higher energies. The last feature, however, is not expected in
continuum models. Indeed, the sum in the definition~(\ref{eq:13}) of
$F_N(E)$ runs over 
all initial states with energy $E$ and multiplicity $N$. These 
include, in particular, the state where  $N-1$ particles
perform transition at smaller energy $E_0$ and one 
spectator
particle carries the energy excess $\Delta E = E - E_0$, see
Fig.~\ref{fig:diagram}. The suppression exponent of the last process
is 
$F_{N-1}(E_0) \approx F_N(E_0)$, where correction of order $g^2 \theta
\ll 1$ is ignored, cf. Eq.~(\ref{eq:16}). Since the inclusive
transition is less suppressed than the exclusive one, we conclude that
$F_N(E) \leq F_N(E_0)$, i.e.\ $F_{N}(E)$ is a non--increasing function of
energy. 

The above argument suggests that the negative values of $T\propto
-\partial_E F_N$ in Fig.~\ref{fig:Tm0}b are lattice artifacts and one
should be careful with the continuum limit $\Delta x \to 0$. Indeed,
Fig.~\ref{fig:Tm0}c shows that the high--frequency modes of solutions with
$E>E_{rt}(N)$ are enhanced as compared to the case of the real--time
instanton and Eq.~(\ref{eq:18}). To perform the quantitative 
comparison, we boldly assume that the semiclassical solutions at $T<0$
have the form $\phi_s(x) = \phi_{rt}(x) + \delta \phi(x)$, where
$\delta \phi$ consists of high--frequency modes evolving
linearly in the real--time instanton background. Ignoring reaction of
the latter on $\delta  \phi$, we write, 
\begin{equation}
\label{eq:26}
\delta \phi(x)  = \int
\frac{dk}{4\pi\omega_k} \left[\delta c_k\, \mathrm{e}^{-ik\cdot x} +
  \delta c_k^*\, \mathrm{e}^{ik\cdot x} \right]\;.
\end{equation}
This solution is real at $t\to +\infty$ because 
 $\delta c_{k}$ and $\delta c_k^*$ are the mutually conjugate
constants. One expresses $\delta c_k$ from the initial condition
(\ref{eq:39}): $\delta c_k = (b_k^{(rt)}\gamma_k - 
a_k^{(rt)})/(1-\gamma_k)$, where $\gamma_k =
\mathrm{e}^{-2\omega_k T   - \theta}$. Here and below we mark
quantities related to the real--time instanton 
with ``$(rt)$'', e.g.\ $a_{k}^{(rt)} = \mathrm{e}^{-\theta}
b_k^{(rt)}$. We obtain the energy  distribution for the solution with $T<0$,
\begin{equation}
\label{eq:24}
\varepsilon_k \equiv \frac{1}{4\pi} (a_k^{(rt)} + \delta c_k )(b_k^{(rt)*} + \delta
c_k^*) =  \varepsilon_k^{(rt)} \, \frac{\sinh^2 (\theta/2)}{\sinh^2 
  (\omega_k T + \theta/2)}\;,
\end{equation}
where $\varepsilon_k^{(rt)} = a_k^{(rt)} b_k^{(rt)*}/4\pi$.
Recall that Eq.~(\ref{eq:24}) is based on a crude assumption that the solutions
with $E>E_{rt}(N)$ differ from the real--time instantons only in 
linearly evolving high--frequency modes. Nevertheless, this scaling
works 
well: in Fig.~\ref{fig:Tm0}c the function (\ref{eq:24}) (points) coincides
with the actual energy distribution (solid line). In
Appendix~\ref{sec:semicl-solut-at} we derive Eq.~(\ref{eq:24})
in somewhat different approach.

Distributions (\ref{eq:24}) are qualitatively different at
$T>0$ and $T<0$. In the former case the high--frequency asymptotics
of $\varepsilon_k$ is consistent with Eq.~(\ref{eq:18}) which implies 
smooth continuum limit. At $T<0$, however, 
the function~(\ref{eq:24}) develops a nonintegrable singularity
at ${\omega_k = \omega_* \equiv -\theta /2T}$. Technically, this feature
is related to the fact that the linear mode with frequency $\omega_*$
satisfies reality conditions at $t\to \pm \infty$, see
Eqs.~(\ref{eq:57}) and (\ref{eq:39}). Solutions can accumulate macroscopic  
energy in this mode before its amplitude $\delta c_{k_*}/\omega_* \sim 
\varepsilon_{k_*}^{1/2}/\omega_*$ becomes large and linear
approximation (\ref{eq:26}) breaks down. Importantly, the latter
energy tends to infinity as $T\to -0$ or $\omega_* \to
+\infty$. 

Lattice solutions do not feel the singularity in 
Eq.~(\ref{eq:24}) if the maximal lattice frequency $\omega_{max} =
2/\Delta x$ is below $\omega_*$. If we decrease $\Delta x$ with
constant $(T,\, \theta)$, the value of $\omega_{max}$ approaches
$\omega_*$ and the solution gains energy. Vice versa, at fixed energy
one obtains smaller $|T|$ at smaller $\Delta x$. If Eq.~(\ref{eq:24})
is valid up to infinitesimally small $\Delta x$,
parameter $T$ approaches zero as $\Delta x\to 0$, so that
$\omega_*$ is kept outside of the lattice frequency range and the
energy remains finite. The respective semiclassical solutions arrive at the
real--time instanton plus modes with infinitely high frequency and
vanishingly small amplitude carrying the energy excess $E -
E_{rt}(N)$. These solutions do not have smooth continuum limits.

\begin{figure}[t]
\hspace{4cm}(a) \hspace{8.5cm}(b)

\includegraphics[width=7.5cm]{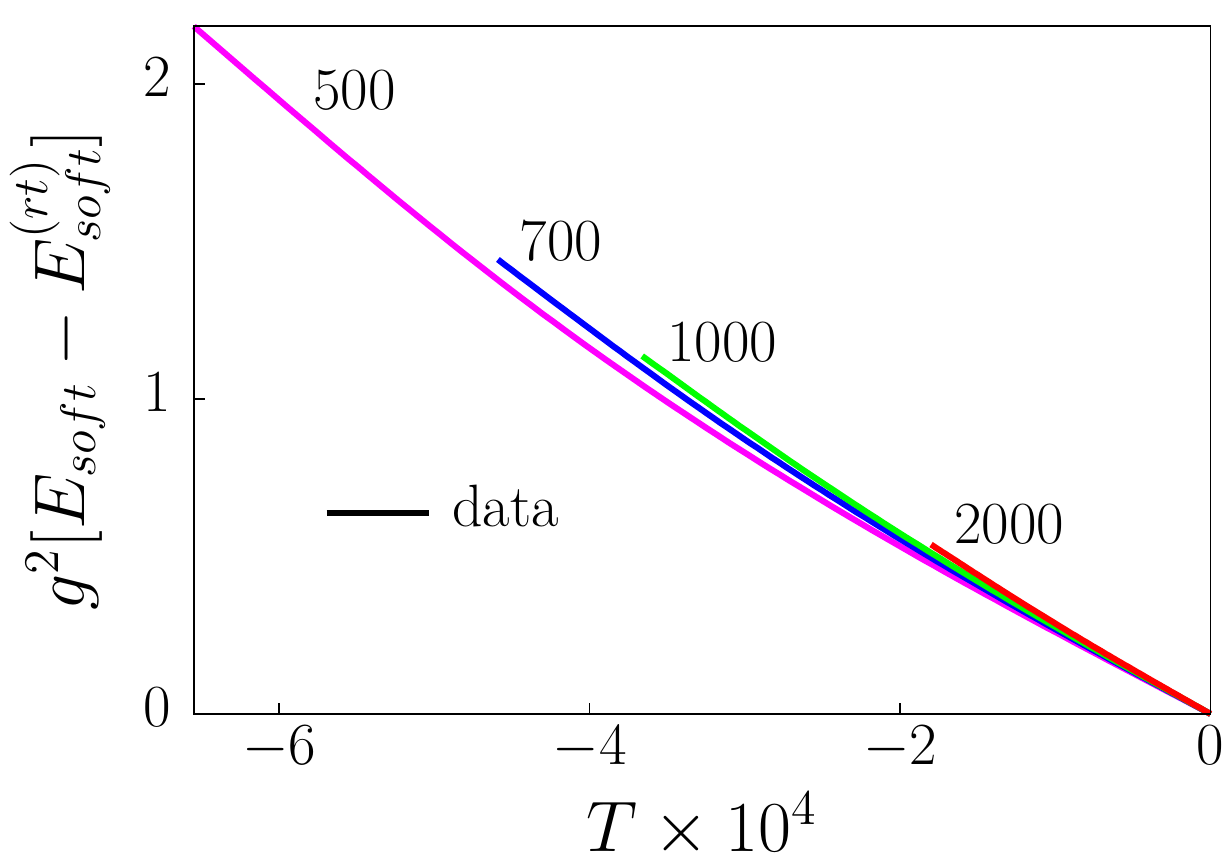}\hspace{1.5cm}
\includegraphics[width=7.5cm]{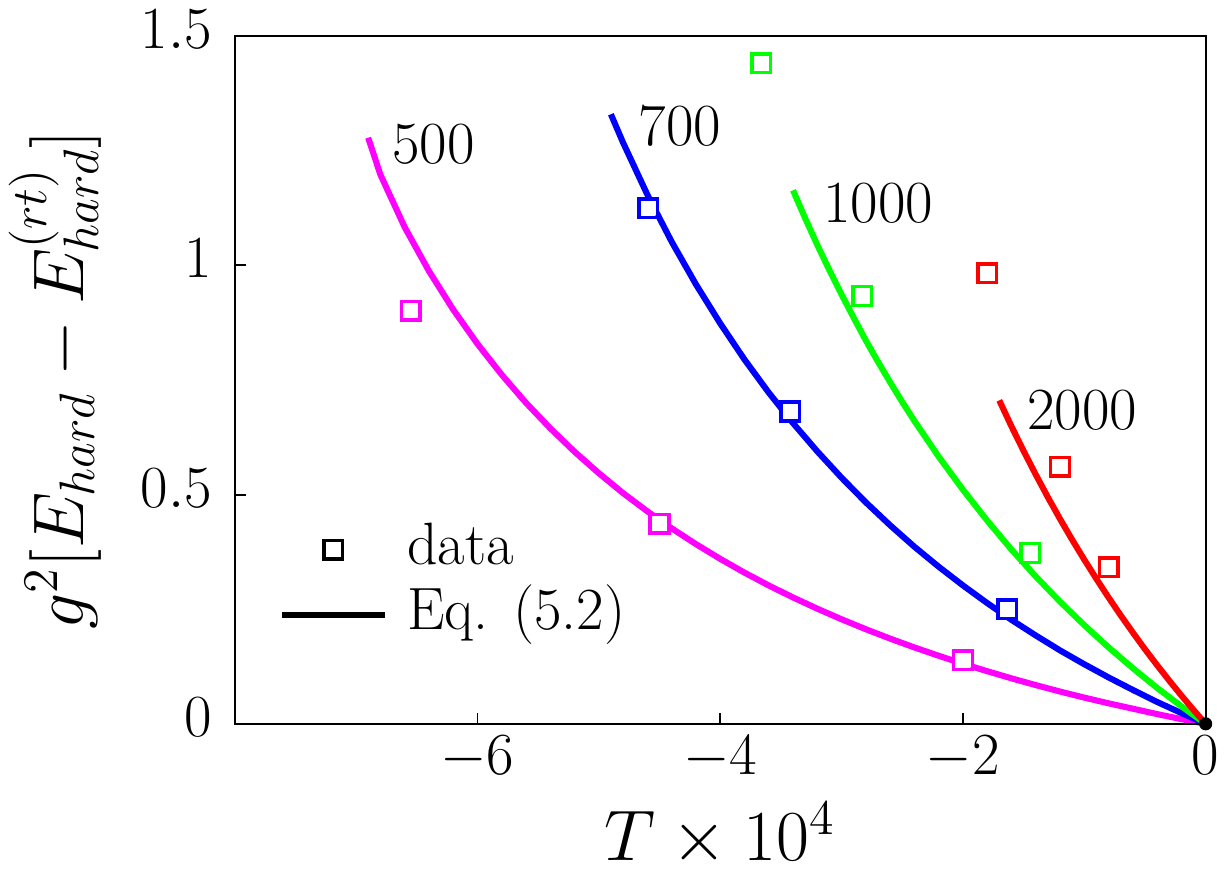}
\caption{Energies of (a) soft and (b) hard modes of solutions at
  $E>E_{rt}(N)$ and different $N_x$ (numbers near the lines);
  $\theta = 0.199$.\label{fig:Esofthard}}
\end{figure}
The scaling property of our solutions is best summarized by dividing
the total energy into ``soft'' and ``hard''
parts, i.e.\ $E = E_{soft} + E_{hard}$, involving modes with $\omega_k
< \omega_\Lambda$ and $\omega_\Lambda 
< \omega_k < \omega_{max}$ in Eq.~(\ref{eq:14}). In
Fig.~\ref{fig:Esofthard} we plot $E_{soft}(T)$ and $E_{hard}(T)$ for
$\omega_\Lambda = 70m$ and several values of $N_x$. Predictably, 
$E_{soft}$ is not sensitive to $N_x$, while $E_{hard}$ sharply depends
on it according to Eq.~(\ref{eq:24}) (lines in
Fig.~\ref{fig:Esofthard}b). Since the data points in
Fig.~\ref{fig:Esofthard}b are well approximated by Eq.~(\ref{eq:24}),
the value of $T$ approaches zero as $N_x \to +\infty$ at fixed
$E_{hard}$, so that $E_{soft} \to E_{soft}^{(rt)}$ in
Fig.~\ref{fig:Esofthard}a. 

A remark is in order. Our numerical results show that at high enough 
$E-E_{rt}(N)$ the singularity $\omega_*$ falls below $\omega_{max}$
and interaction of high--frequency modes becomes important. We find, 
however, that the above qualitative picture holds even in this
case. In particular, $|T|$ decreases with $N_x$ at fixed total energy
$E$, see 
Fig.~\ref{fig:Tm0}b. We therefore expect that the respective continuum
limit has the same properties: parameter $T$ vanishes as $N_x \to
\infty$ and lattice solutions approach the real--time instanton plus
high--frequency modes.

It is important to point out that the lattice solutions lacking
smooth continuum limits at $E>E_{rt}(N)$, are still capable of
describing tunneling transitions in that region. Namely, instead of
discretizing the semiclassical equations \eqref{eq:54}
one 
can start from the lattice path integral for the
cross section~(\ref{eq:13}). The latter is an ordinary integral over
$N_t \times 2N_x$ variables $\phi_{ij}$. At $g^2 \to 0$ it can be
evaluated in the saddle--point approximation, where the saddle--point values of 
$\phi_{ij}$ are the semiclassical lattice solutions. In this
approach the existence of well--defined continuum limits of the
latter solutions is irrelevant. The overall semiclassical results
are reliable if the exponent $F_{N}(E)$  has a limit  at
$\Delta x \to 0$. Then the lattice solutions point at the dominant
mechanism of quantum transition.

We have already argued that our lattice solutions with
$E>E_{rt}(N)$ describe processes shown in Fig.~\ref{fig:diagram}:
they tend to the real--time instanton with energy $E_{rt}(N)$ plus
a few spectator particles in the form of 
high--frequency waves carrying the energy excess $E - E_{rt}(N)$. The respective
suppression exponent is constant because $T \propto -\partial_E F_N$
approaches zero as $\Delta x \to 0$. We finally conclude that $F_N(E)$
is constant and equal to $F_{min}(N)$ at energies above the threshold
$E_{rt}(N)$, see 
Fig.~\ref{fig:potential_energy}c and cf. Refs.~\cite{Voloshin:1993ks,
  Levkov:2004ij}.

\section{Stability of perturbative expansion around the real--time $\mbox{instanton}$}
\label{sec:pert-expans-at}
Since the inclusive cross section does not grow exponentially with energy
at $E>E_{rt}$, one assumes that the terms of its perturbative
expansion in $g^2$ around the real--time instanton are also  bounded as
$E\to +\infty$. Then one can compute the collision--induced amplitudes
and cross sections perturbatively at $N=2$ and arbitrary high energies.

To test this property, we directly address
the two--particle inclusive cross section of collision--induced false
vacuum decay, 
\begin{equation}
\label{eq:41}
\sigma(E) = \sum_{\Psi_f}\left|\langle \Psi_f |
\hat{U}(t_f,\, t_i) \hat{a}^{\dag}_{p_2}\,
\hat{a}^{\dag}_{p_1}|\Psi_0
\rangle \right|^2\;,
\end{equation}
where $t_{i,f} \to \mp \infty$, the initial state describes two particles with total
momentum $P \equiv p_1 + p_2 = (E, \, 0)$ in the false vacuum 
$\Psi_0$, each final state $\Psi_f$ contains an expanding bubble of true
vacuum, and we ignore prefactors. Let us check the possibility of
extracting the cross section (\ref{eq:41}) from the perturbative
Green's functions in the backgrounds of the real--time instantons, so
that the corrections to the perturbative calculations remain
small as $E\to +\infty$.

In this Section we compute only the ``factorized'' contribution
to $\sigma(E)$ which was dominant and responsible for the exponential
growth of the inclusive cross section in the Euclidean approach of
Sec.~\ref{sec:grow-ampl-backgr}. We will see that this
contribution is exponentially suppressed if the real--time instanton
is used as a background. A consistent perturbative expansion around
$\phi_{rt}(x)$ will be developed in the next Section.  

We start from the Green's function (\ref{eq:36}) between the false vacuum $\Psi_0$
and the dominant final state $\Psi_{rt}$ of the real--time
instanton $\phi_{rt}(x)$. In the leading order of the perturbation theory
we substitute $\phi$ with $\phi_{rt}(x-x_0)$ in the integrand of 
Eq.~(\ref{eq:36}) and find,
\begin{equation}
\label{eq:30}
{\cal G}_{rt} = {\cal A}_{rt}
 \int d^2 x_0 \, \Psi_{rt}^{*}[\phi_{rt}] \; \phi_{rt}(x_1 - 
x_0) \dots \phi_{rt}(x_{n+2}-x_0)\;,
\end{equation}
where\footnote{Recall that $S$ and $\Psi_0$ are independent of $x_0$
  in the limit $t_{i,f} \to \mp\infty$, as we drop the terms
  oscillating with $t_i$ and $t_f$.} 
\begin{equation}
\label{eq:42}
{\cal A}_{rt} = \mathrm{e}^{iS[\phi_{rt}]}\,
\Psi_0[\phi_{rt}]\;.
\end{equation}
Note that the position $x_0$ of $\phi_{rt}$ is not fixed by the
semiclassical equations \eqref{eq:54}. Nevertheless, the
final--state wave functional $\Psi_{rt}[\phi_{rt}]$ depends on
$x_0$ via $\phi_{rt}(x-x_0)$.

We use the LSZ reduction formula and turn Eq.~(\ref{eq:30}) into 
the $2\to n + \Psi_{rt}$ amplitude. Considering the initial particles, we trade two
of $\phi$'s in the integrand for their positive--frequency
residues  $b_{p}^* \mathrm{e}^{-ip\cdot x_0}$, see
Eq.~(\ref{eq:17}). The case of the final--state particles is more
involved because at $t\to +\infty$ the configuration $\phi_{rt}$ contains,
apart from the outgoing waves, an interacting bubble. To handle
this difficulty, we assume that the 
false vacuum decay occurs in a finite volume $|x| < L$ with periodic
boundary conditions. Then at $t\to 
+\infty$ the bubble fills all space and $\phi_{rt}(x)$ reduces to
waves in the true vacuum, 
\begin{equation}
\label{eq:34}
\phi_{rt}(x - x_0) \to \phi_+ + \int \frac{dk}{4\pi \omega_k^{(+)}} \left[c_k
  \mathrm{e}^{-ik \cdot (x-x_0)}  + c_k^* \mathrm{e}^{ik \cdot (x-x_0)}\right]
\qquad \mbox{as} \qquad t \to +\infty\;,
\end{equation}
where $k^\mu = (\omega_k^{(+)},\, k)$ are the on--shell momenta in the
vacuum $\phi_+$. The LSZ formula substitutes $\phi_{rt}(x-x_0)$ with its residue $c_{k}
\mathrm{e}^{i k\cdot   x_0}$ for every final--state particle. 
For clarity below we consider finite large $L$ and do not study the limit
$L \to +\infty$. 

We obtain, 
\begin{equation}
\label{eq:28}
{\cal A}_{2\to n + \Psi_{rt}}' = {\cal A}_{rt} \int d^2 x_0
\; \frac{b_{p_1}^*b_{p_2}^*}{g^{2+n}} \; c_{k_1} \dots c_{k_n}\;
\Psi_{rt}^{*}[\phi_{rt}] \; \mathrm{e}^{ix_0\cdot (k_1 + \dots  +k_n
  - p_1- p_2)}\;.
\end{equation}
Here $p_j$ and $k_i$ are the momenta of the initial and final
particles, the prime of ${\cal A}'$ reminds that only the factorized
contribution is considered. Unlike in Sec.~\ref{sec:grow-ampl-backgr},
we do not explicitly integrate over $x_0$ because $\Psi_{rt}$
depends on it in a nontrivial way. Indeed, since $\phi_{rt}(x)$ is real
at $t\to +\infty$, its dominant final state is a coherent
one~\cite{Khlebnikov:1991th}\cite{Rubakov:1992gi}: in the
interaction representation 
\begin{equation}
\label{eq:27}
|\Psi_{rt}\rangle = \exp\left\{\int dk \,  \frac{ c_k\,
  \hat{c}_k^\dag}{4\pi g^2 \omega_k^{(+)}}
\right\}  |\Psi_+\rangle\;, 
\end{equation}
where we introduced creation operators $\hat{c}_k^\dag$ in the true
vacuum $\Psi_+$. Parameters $c_k$ of $\Psi_{rt}$ are precisely the
final--state residues $c_k$ in Eq.~(\ref{eq:34}).  
One can extract dependence of the wave functional $\Psi_{rt}[\phi_{rt}]$
on $x_0$ from its transformation properties $\hat{c}_k^\dag \to
\hat{c}_k^{\dag} \mathrm{e}^{-ik\cdot x_0}$ and 
$\Psi_+ \to \mathrm{e}^{-i t_0 E_+} \Psi_+$ under spacetime shifts
$x\to x+x_0$, where $E_+ = 2LV(\phi_+) < 0$ is the energy of the true
vacuum. 

The amplitude~(\ref{eq:28}) has almost factorized form. In
Appendix~\ref{sec:mult-phase-volume} we integrate over the
phase--space volume and obtain the inclusive
cross section,
\begin{equation}
\label{eq:29}
\sigma'(E) = |{\cal A}_{rt}\;b_{p_1}b_{p_2} \Psi_+[\phi_{rt}]|^2  \int
d^2 \lambda \, \exp\left\{iP \cdot \lambda -iE_+ \lambda^0 + \int
\frac{dk\, |c_k|^2}{4\pi g^2 \omega_k^{(+)}} \; \mathrm{e}^{-ik\cdot
  \lambda} \right\}\;,
\end{equation}
where $P \equiv p_1+p_2 = (E,\, 0)$. Equation
(\ref{eq:29}) looks similar to Eq.~(\ref{eq:10}) of
Sec.~\ref{sec:grow-ampl-backgr}, and yet it is  entirely
different. First, the initial--state factor $|b_{p_1} b_{p_2}|^2
\propto \mathrm{e}^{-2 E T_*}$ decays exponentially with energy
because higher momentum transfer from the initial particles to the
soft background is less probable, see Eq.~(\ref{eq:18}). Second, the
final--state contribution is nontrivial because $\Psi_{rt}$ is not
an eigenstate of energy, it absorbs different energies in different
cases. Thus, the simple picture of converting all energy into the
multiparticle final states with huge phase volume is lost.

Since the exponent in Eq.~(\ref{eq:29}) is large, the integral over
$\lambda$ is evaluated in the saddle--point approximation. One finds
the extremum of the exponent $\lambda_s^\mu = (-2i  T',\, 0)$  satisfying 
\begin{equation}
\label{eq:31}
E = E_+ + \int dk \, \frac{|c_k|^2}{4\pi g^2} \; \mathrm{e}^{-2
  T'\omega_k^{(+)}}\;.
\end{equation}
At $E= E_{rt}$ the solution is $T' = 0$ because the
right--hand side of Eq.~(\ref{eq:31}) coincides with
the final energy 
of the real--time instanton, cf. Eq.~(\ref{eq:34}). At $E > E_{rt}$ 
one obtains $T' < 0$. Substituting $\lambda_s$ into
Eq.~(\ref{eq:31}), we find,  
\begin{equation}
\label{eq:32}
\sigma'(E) = |{\cal A}_{rt} \,\Psi_+[\phi_{rt}]|^2 \exp\left\{ 2E(T'-T_*) -2E_+ T'
  + \int
\frac{dk\, |c_k|^2 \, \mathrm{e}^{-2T'\omega_k^{(+)}}}{4\pi g^2 \omega_k^{(+)}
  }\right\} = \mathrm{e}^{-F'(E)/g^2}\,,
\end{equation}
where the suppression exponent is introduced in the second
equality. 

Equation~(\ref{eq:31}) implies that the suppression exponent 
of $\sigma'(E)$ grows at $E>E_{rt}$,
\begin{equation}
\label{eq:47}
\frac{dF'}{dE} = 2g^2(T_*-T') > 0\;.
\end{equation}
In the next Section we will demonstrate that at $E = E_{rt}$ the
exponent $F'(E)$ coincides with the true semiclassical exponent
$F(E)$. The latter, however, is constant at high energies. Thus, the
factorized contribution \eqref{eq:32} is exponentially subdominant,
$F'>F$, at $E>E_{rt}$.

\section{Perturbative method at high energies}
\label{sec:pert-descr-at}
Before considering the dominant contribution, we remark that the
perturbative approach of the previous Section is incomplete. Indeed,
there is a large family of real--time instantons
parametrized\footnote{Recall that we work in the finite--$L$ box which
  explicitly breaks Lorentz symmetry. Boosted real--time instantons
  should be taken into account in the infinite--volume limit.} 
with $N=N_0$ and position $y_0$. Every solution from this
family has its own dominant final state $\Psi_{rt}$ which also depends
on $N_0$ and $y_0$, cf. Eq.~(\ref{eq:27}). In this Section we change
notation $N\to N_0$ for the real--time instanton parameter, to
distinguish it from the number of colliding particles in the process;
from now on, the latter equals two. We stress that the values of $N_0$
and $y_0$ characterize the background for the perturbative expansion.
Their values are selected to achieve better convergence. 

Our receipt for perturbative evaluation of the collision--induced cross
section at high energies and two initial particles is summarized as
follows~\cite{Rubakov-conjecture}. One starts from  
the $(n+2)$--point Green's function (\ref{eq:36}) between the false vacuum $\Psi_0$
and the dominant final state $\Psi_{rt}$ of the real--time
instanton  $\phi_{rt}^{(N_0)}(x-y_0)$, cf. Eq.~(\ref{eq:27}). One
evaluates the path integral for the Green's function perturbatively in
the background of $\phi_{rt}$: substitutes $\phi(x) =
\phi_{rt}^{(N_0)}(x-x_0) + g\delta\phi(x)$ and expands the integrand in
$g\delta \phi$. Integral over the would--be flat direction $x_0 \ne y_0$
remains in the Green's function, cf. Eq.~\eqref{eq:30}. One finally extracts
the perturbative amplitudes ${\cal A}_{2\to n + \Psi_{rt}}$ from
the LSZ formula,  turns them into the inclusive cross section with the
machinery of Appendix \ref{sec:mult-phase-volume}, and
takes the limit $N_0\to 0$. The result  is a perturbative expansion for
$\sigma(E)$ which, as we are going to argue, is applicable at
arbitrary high energies.

\begin{figure}[t]
\vspace{8mm}

\hspace{4.2cm} (a) \hspace{7.8cm} (b)

\vspace{-10mm}
\centerline{\includegraphics[width=7.5cm]{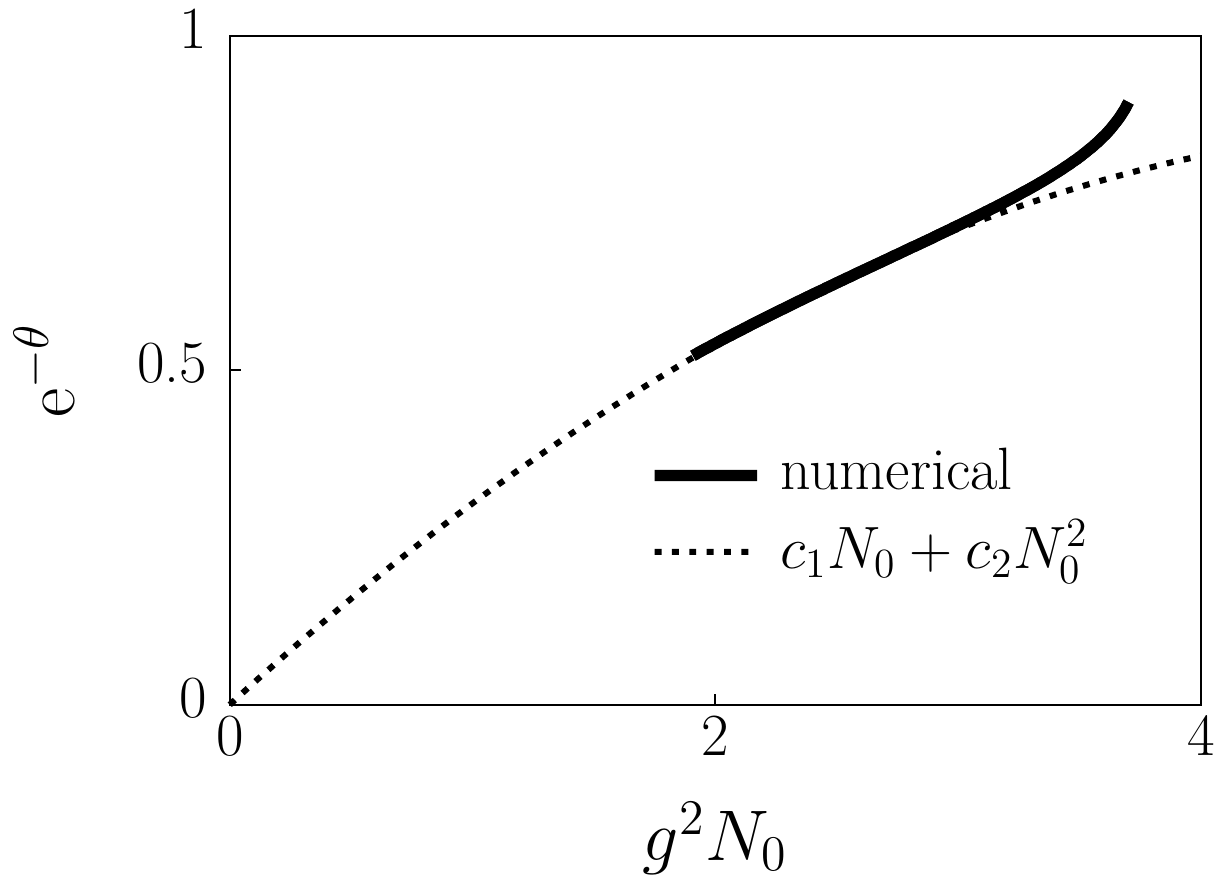}\hspace{1cm}
\includegraphics[width=7.5cm]{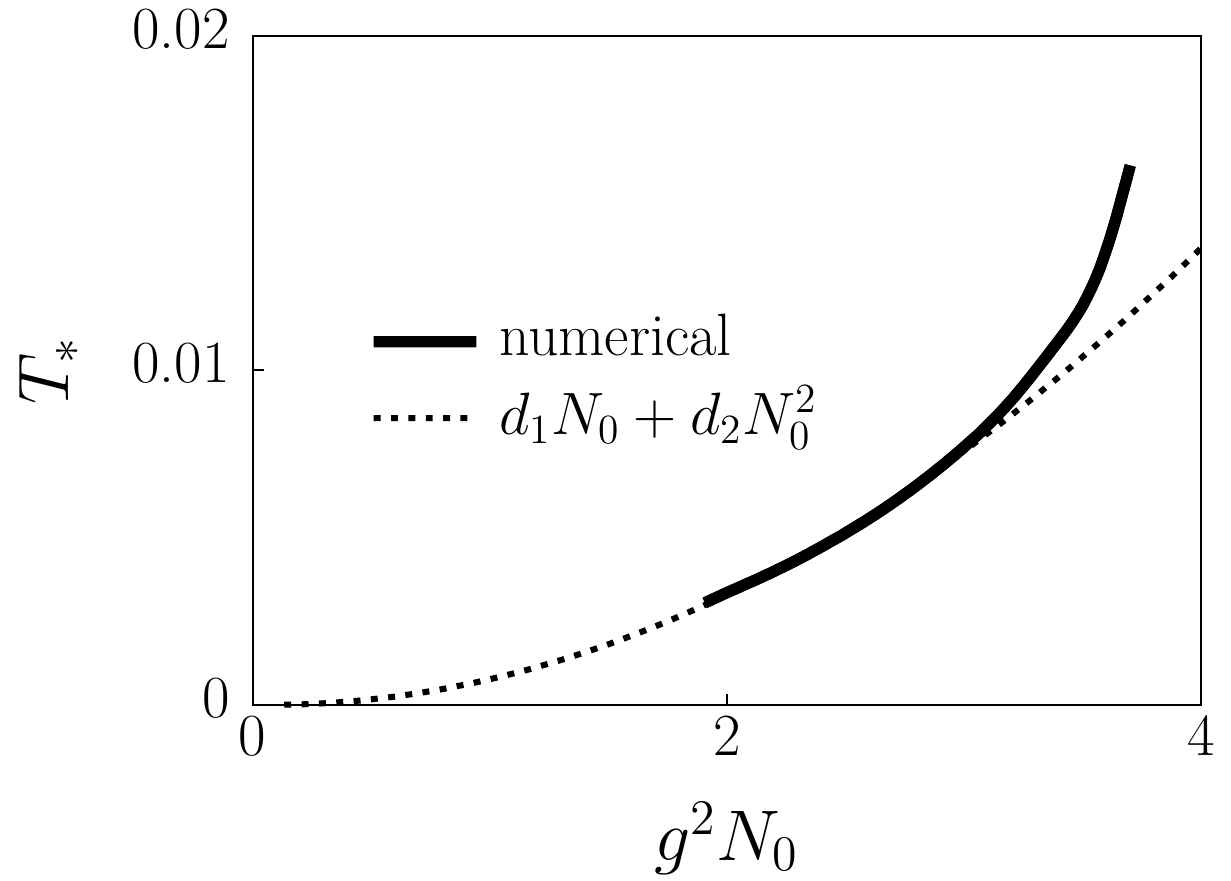}}
\caption{Quantities (a) $\mathrm{e}^{-\theta}$ and (b) $T_*$
  characterizing   the real--time instanton as functions of its
  parameter $N_0$. Numerical data at 
  $N_t\times N_x=11000\times 3000$ (solid lines) are fitted
  with quadratic polynomials in $N_0$ (dotted
  lines). \label{fig:ethetart}} 
\end{figure}
Let us explain the role of the auxiliary parameter $N_0$
characterizing the background solution. On the one hand, recall that the real--time instanton
$\phi_{rt}^{(N_0)}$ describes transition from the initial states with
$N_0$ particles: the limit $N\equiv N_0\to 0$ formally corresponds to 
the vacuum initial state. Indeed, at
$\theta\to +\infty$ one simultaneously obtains Feynman initial
condition $a_k \to 0$ for $\phi_{rt}^{(N_0)}$ and $N_0\to 0$, see
Eqs.~(\ref{eq:39}) and (\ref{eq:14}). Thus, the real--time instanton
with $N_0=0$ and $x_0 = y_0$ is the
formal saddle--point configuration\footnote{Recall that 
  $\phi_{rt}$  extremizes the  classical action
  and, by construction of $\Psi_{rt}$,  at $x_0 =
  y_0$ serves as the saddle--point
  configuration for the  integral with the final state in
  Eq.~(\ref{eq:36}). Note also that the two initial particles of the process are
  represented by the two $\phi$--factors in the integrand which do not
  change the initial saddle--point conditions $a_k = 0$. The latter
  are satisfied by $\phi_{rt}^{(N_0)}$ at $N_0 \to 0$.} for 
the path integral (\ref{eq:36}), and the
perturbative series around it constitute the ordinary
saddle--point expansion. On the other hand, the energy $E_{rt}(N_0)$ of
the real--time instanton represents the extremum, $\partial_E F_N=0$, and 
stays nonzero in the limit $N_0\to 0$, cf. Fig.~\ref{fig:Ert}b. This
means that $\phi_{rt}^{(N_0)}$ is singular at $N_0=0$ because its
typical frequencies $\omega_k \sim E/N_0$ are infinite. We therefore
develop perturbative expansion around the smooth configurations with $N_0>0$
and send $N_0\to 0$ in the end of calculations.

In Fig.~\ref{fig:ethetart} we plot parameter $\mathrm{e}^{-\theta}$ of
the real--time instanton and distance $T_*$ to its closest singularity
as functions of $N_0$. Numerical data (solid lines) are well fitted by
quadratic polynomials with zeros at $N_0=0$ (dashed lines). The graphs
support our expectation that as $N_0\to 0$, the real--time instanton
$\phi_{rt}^{(N_0)}$ tends to 
a singular configuration with vacuum initial conditions.

Expanding the integrand of Eq.~(\ref{eq:36}) in $g\delta \phi$, we
obtain Feynman rules involving points, propagators and vertices,
\begin{equation}
\label{eq:35}
\includegraphics[width=7mm]{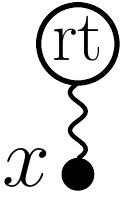}
= \frac{\phi_{rt}}{g}\Big|_{x-x_0}\;, \;\;\;
\includegraphics[width=1.67cm]{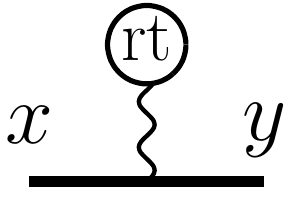}
= \langle \delta \phi (x) \delta \phi(y) \rangle_{rt}\;,\;\;\;
\begin{minipage}{2.4cm}
\vspace{-3.5mm}
\includegraphics[width=2.4cm]{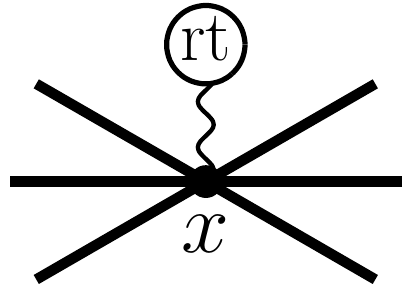}
\end{minipage}
 = g^{m-2} V^{(m)}(\phi_{rt})\Big|_{x-x_0}\;,
\end{equation}
which explicitly depend on $N_0$ and $x_0$. At nonzero $N_0$ or $x_0 \ne
y_0$ one also obtains the tadpoles, i.e.\ terms in action proportional
to $\delta \phi$, which characterize deviation of the background
solution  from the true saddle--point configuration. The tadpoles
coming from the terms at $t\to -\infty$ 
vanish as $N_0\to 0$, and we do not consider them in what
follows. The final--state tadpoles are related to the fact that
$\Psi_{rt}$ is the dominant 
final state for the configuration $\phi_{rt}(x-y_0)$ which is
different from our background $\phi_{rt}(x-x_0)$. We will discuss them 
in the end of this Section. Since the elements in  
Eq.~(\ref{eq:35}) explicitly depend on $x$, energy and momentum are
not conserved along the lines and in the vertices. Rather, the Feynman
rules in momentum space involve structure functions depending on the
momentum $Q$ 
transferred to the background. We will see shortly that the real--time
instanton consumes total energy $Q^0 \approx E_{rt}$ from the initial
particles.   

\begin{figure}

\centerline{(a)\hspace{3.2cm}(b)\hspace{3.1cm}(c)\hspace{3.2cm}(d)\hspace{2mm}}

\vspace{2mm}
\centerline{
\begin{minipage}{3.2cm}
\includegraphics[height=3.2cm]{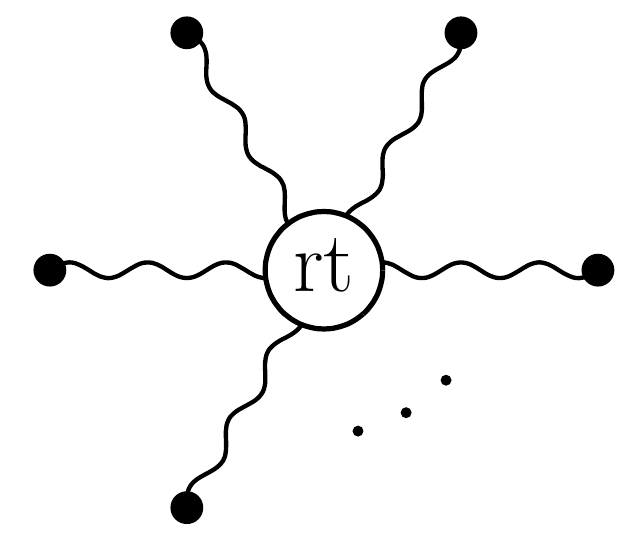}
\end{minipage}
\hspace{1cm}
\begin{minipage}{11cm}
\includegraphics[height=3cm]{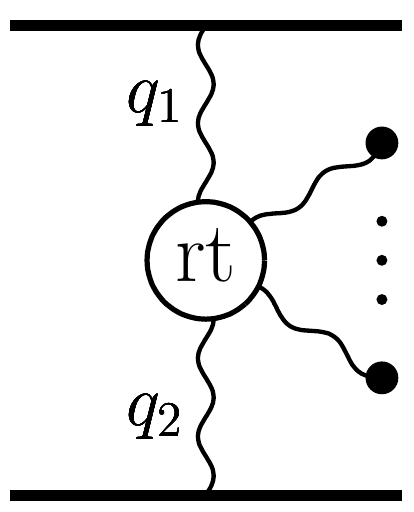}
\hspace{1cm}
\includegraphics[height=3cm]{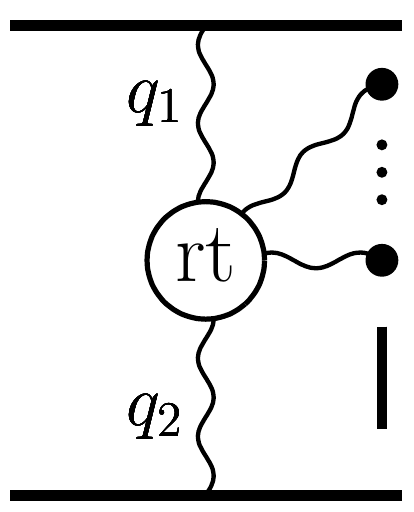}
\hspace{1cm}
\includegraphics[height=3cm]{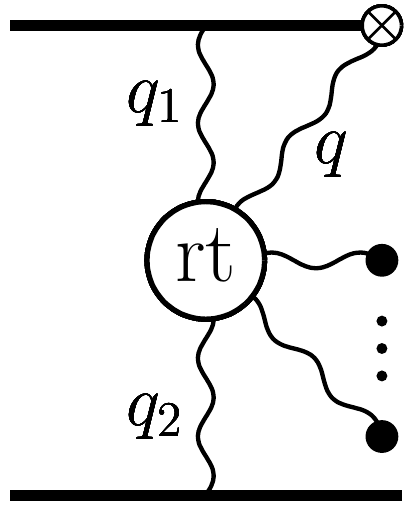}
\end{minipage}}
\vspace{3mm}
\caption{Diagrams for perturbative expansion around the real--time
  instanton.\label{fig:diagrams}}
\end{figure}

At zeroth order of the perturbative expansion one uses $\phi(x) =
\phi_{rt}^{(N_0)}(x-x_0)$ in Eq.~(\ref{eq:36}) and obtains the diagram in 
Fig.~\ref{fig:diagrams}a. Extracting the cross section and summing over
the final states, one arrives at the contribution (\ref{eq:32}) which
is exponentially subdominant. Indeed, we argued in Eq.~(\ref{eq:47}) that
the suppression exponent $F'(E)$ of this contribution
grows with energy at $E>E_{rt}$. Besides,  at $E = E_{rt}$ and $N_0 \to 0$ we
have $T' = T_* = 0$ and therefore
\begin{equation}
\label{eq:48}
F'(E_{rt}) = 2g^2\mathrm{Im}\, S[\phi_{rt}] - g^2\ln |\Psi_0|^2 - g^2\ln
|\Psi_+|^2 - \int \frac{dk\,|c_k|^2}{4\pi \omega_k^{(+)}} = F(E_{rt})
+ O(N_0)\;.
\end{equation}
Here we used Eqs.~(\ref{eq:42}) and (\ref{eq:15}), vacuum wave
functionals 
\[
\Psi_0[\phi] = \exp\left\{-\int \frac{dk \, \omega_k}{4\pi g^2} \,
\phi(k) \phi(-k)\right\}_{t_i}
\] 
and $\Psi_+[\phi]$, spatial Fourier transform $\phi(k)$ of
configuration $\phi$, and representations (\ref{eq:17}), 
(\ref{eq:34}) of the real--time instanton\footnote{Recall also that
  $F_N(E) =  
  F(E) - g^2\theta N + O(N)$ as we argued in the end of 
  Sec.~\ref{sec:from-euclidean-real}.}.  Since
    the dominant exponent $F(E)$ is constant 
    at high energies, we repeat that $F'(E)>F(E)$ at $E>E_{rt}$,
    i.e.\ the factorized diagram in Fig.~\ref{fig:diagrams}a is
    negligible.\footnote{One finds that as $L \to \infty$, the energy
      of the true vacuum becomes infinite and therefore $T' \to
      0$. This means that in the true infinite--volume limit (not
      considered here) the contribution
      (\ref{eq:32}) becomes comparable to the dominant one.} 

The dominant diagram at $E>E_{rt}$ is shown in
Fig.~\ref{fig:diagrams}b. It describes propagation of two initial
particles which transfer momentum $Q = q_1 + q_2$ to the
background. The respective transition amplitude
is\footnote{Symmetrization with respect to permutations of $k_1,\,
  \dots,\, k_n$ is assumed.},
\begin{equation}
\label{eq:33}
{\cal A}_{2\to n + \Psi_{rt}} = {\cal A}_{rt} \, n(n-1) \;\int d^2 x_0\, 
D_{p_{1},\, k_1}\, D_{p_2,\, k_2} 
\; \frac{c_{k_3} \dots c_{k_n}}{g^{n-2}}\;
\Psi_{rt}^{*}[\phi_{rt}] \; \mathrm{e}^{ix_0\cdot (k_3 + \dots  + k_n
  - q_1 - q_2)}\;,
\end{equation}
cf.\ Eq.~(\ref{eq:28}). Here $q_i \equiv p_i - k_i$ and
$D_{p,\, k} \cdot\mathrm{e}^{-iq \cdot   x_0}$ is the double
residue of
the propagator $\langle 
\delta \phi\, \delta \phi\rangle_{rt}$. In Appendix
\ref{sec:mult-phase-volume} we convert the amplitude into the inclusive cross
section,
\begin{multline}
\sigma(E) = |{\cal A}_{rt}\;\Psi_+[\phi_{rt}]|^2  \int 
  \frac{d k_1 d k_2}{32 \pi^2 \omega_{k_1}^{(+)} \omega_{k_2}^{(+)}} \;\;
  |D_{p_1,\, k_1} \, D_{p_2,\, k_2} + D_{p_1,\, k_2} \, 
  D_{p_2,\, k_1}|^2 \\\times  \int  
d^2 \lambda \,  \exp\left\{iQ \cdot \lambda -iE_+ \lambda^0 + \int
\frac{dk\, |c_k|^2}{4\pi g^2 \omega_k^{(+)}} \; \mathrm{e}^{-ik\cdot
  \lambda} \right\}\;,
\label{eq:53}
\end{multline}
where $\lambda \equiv y_0 - x_0$ and we omit trivial prefactors. The
first line in this expression is the naive Feynman diagram in the
background of the real--time instanton. The factor in the second line
is related to 
the on--shell final state of the background process. It is the 
same as in Eq.~(\ref{eq:29}). In Sec.~\ref{sec:pert-expans-at} we
demonstrated that this factor is 
sharply peaked around $Q^\mu = (E_{rt}, \, 0)$ with fluctuations of
order $\Delta Q \sim gE_{rt}$.

Evaluating the integrals over $Q$ and $\lambda$ in the saddle--point
approximation, we get with exponential precision,
\[
\sigma(E) =
\mathrm{e}^{-F(E_{rt})/g^2} \qquad \mbox{at} \qquad E>E_{rt}\;,
\]
see Eqs.~(\ref{eq:32}) and~(\ref{eq:48}). We thus obtained constant
suppression exponent $F=F(E_{rt})$ at $E>E_{rt}$ which was deduced in
Sec.~\ref{sec:transitions-at-ee_rt} on the basis of sophisticated
numerical analysis. Note that our perturbative method can be applied
for calculating the prefactor: one just has to estimate the
energy--independent saddle--point determinant in the Green's  
function~(\ref{eq:36}) and collect few simple prefactors 
in the above calculations.

The first perturbative correction to the dominant contribution is
shown in Fig.~\ref{fig:diagrams}c. It involves the same relative
factor $g^2 n^2$ as in the low--energy calculation of
Sec.~\ref{sec:grow-ampl-backgr}. This time, however, the number $n$ of
the point--like factors $\phi_{rt}$ representing the particles added to
the background, is relatively small and does not grow with the collision
energy $E$. Indeed, any energy transfer to the background above $Q =
E_{rt}$ is cut off by the final--state factor. This means that 
${n \sim  |Q-E_{rt}|/m \sim g^{-1}}$, and the diagram
in Fig.~\ref{fig:diagrams}c is of the same order as the dominant
one. Importantly, it does not grow with energy. Resummation of these
disconnected diagrams~\cite{Arnold:1990va,Khlebnikov:1990ue} leaves us
with corrections involving vertices which are suppressed by the
true expansion parameter $g$. 

Let us finally discuss the tadpoles. We saw that the 
parameter $y_0$ of $\Psi_{rt}$ does  not coincide with the
position $x_0$ of the background solution: the integral over the
difference $\lambda \equiv y_0 - x_0$ enters Eq.~(\ref{eq:53}). Thus, 
$\phi_{rt}(x-x_0)$ is not the true saddle--point
configuration of the integral~(\ref{eq:36}), perturbative
expansion around it starts from the linear term in $\delta 
\phi$. We obtain the tadpole 
\begin{equation}
\label{eq:50}
\begin{minipage}{1.45cm}
\includegraphics[width=1.45cm]{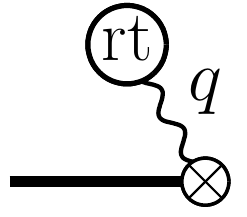}

\vspace{4mm}
\end{minipage}
\hspace{2mm}
\begin{minipage}{5cm}

\vspace{5mm}
$= \displaystyle\frac{c_q^*\, \mathrm{e}^{-iq\cdot x_0}}{4\pi g
  \omega_k^{(+)}} \, \left(\mathrm{e}^{-iq\cdot \lambda} - 1\right)$
 \end{minipage}
\end{equation}
which should be integrated over $q$ with the final--state residue of
the propagator. Apart from the additional integration, this
tadpole is similar to the final on--shell particle in the
amplitude. One may think that the dominant contribution includes diagrams
with the tadpoles attached to the hard propagators, 
e.g.\ Fig.~\ref{fig:diagrams}d. However, all momentum $p_1 = q_1+q$ of
the propagator with the tadpole is transferred to the real--time
instanton. Indeed,  Eq.~\eqref{eq:50} is proportional to the exponent
$e^{-iq\cdot x_0}$ which accumulates transferred momentum in the amplitude,
cf. Eq.~(\ref{eq:33}). Then the total transferred energy in the
process in Fig.~\ref{fig:diagrams}d is higher than $E_{rt}$ and the
respective contribution to the cross section is  exponentially small.

We see that the tadpoles infest the final--state propagators in
Fig.~\ref{fig:diagrams}c and 
subdominant diagrams\footnote{One can check that the tedpoles do not
  change our conclusion about exponential suppression of the
  factorized contribution.}. Note, however,  that the element
(\ref{eq:50}) is $O(g^{0})$ 
at best. Indeed, at small $q$ the bracket in Eq.~\eqref{eq:50} is
proportional to $g$ because $\lambda \sim g$ at $E\approx E_{rt}$.  At
$q \sim m/g$ the tadpole is suppressed by 
the factor $c_{q}^*/\omega_q^{(+)}$ which vanishes at high $q$ because the
energy of the real--time instanton is finite. Thus, the
tadpoles also do not break the perturbative expansion.

We conclude that the perturbative expansion around the real--time
instanton is reliable at arbitrarily high energies.

\section{Summary and discussion}
\label{sec:conclusions}
In this paper we studied collision--induced tunneling  in
field theory. We paid special attention to the case of two colliding 
particles with high total energy $E$.  As a
playground we considered induced false vacuum decay in $(1+1)$
dimensions. We demonstrated that the suppression exponent $F_N(E)$ of
this process decreases with energy, reaches minimum $F = F_{min}(N)$ at $E =
E_{rt}(N)$ and remains constant at higher energies.

Our methods rely on existence of the real--time instantons --- a
special class of semiclassical solutions describing inclusive
collision--induced  transitions from the initial states with $N$ particles and
arbitrary energies. The minimal suppression $F_{min}(N)$ and threshold 
energy $E_{rt}(N)$ are computed as functionals on these
solutions. Real--time instantons were first observed in the toy model
of Ref.~\cite{Levkov:2004ij}. Here we numerically obtained them in the case 
of $(1+1)$--dimensional false vacuum decay. We expect that these
solutions exist for other collision--induced tunneling processes. One
can verify this expectation on a case--to--case basis by solving the
respective semiclassical boundary value problem in a given model. 

Importantly, we argue on general grounds that the real--time
instantons are complex solutions evolving in real 
time (hence the title). This property is very unusual for the semiclassical
solutions related to exponentially suppressed transitions, it leads to
far--reaching consequences. We find that scattering of high--energy quantum 
particles in the backgrounds of the real--time instantons resembles
scattering in vacuum because energy exchange between the
particles and  the soft background occurs with
exponentially small probability. This situation is radically different from that at low
energies where the Euclidean semiclassical solutions recycle any
additional energy into exponentially large probability
factors. Starting from the real--time instantons, we develop a 
perturbative description of the two--particle collision--induced processes 
at high energies. We 
demonstrate that this description remains valid
at arbitrary high energies. Our method shows that the suppression
exponent $F(E)\equiv F_{2}(E)$ is constant at
$E>E_{rt}(2)$. The collision--induced 
transitions in this regime  involve transfer fixed energy $E_{rt}$ from the two colliding particles to the soft background;
the energy excess $E - E_{rt}$ remains in the initial particles
till the end of the process.  Note that our perturbative methods can
be easily generalized for calculating  prefactors or exclusive
cross sections.

We conclude that the real--time instantons, if exist for a given
collision--induced process, provide powerful
perturbative framework and guarantee constant suppression exponent 
$F(E)=F_{min}(2)$ of this process at energies above a certain threshold $E_{rt}(2)$.

\acknowledgments
We are grateful to S.M. Sibiryakov and F.L. Bezrukov for motivation and
to V.A. Rubakov~\cite{Rubakov_60} for criticism. This work is
supported by the RSCF grant 14-22-00161. D.L. thanks EPFL for hospitality.

\appendix
\section{Multiparticle cross sections}
\label{sec:mult-phase-volume}
In this Appendix we evaluate inclusive cross sections for the
amplitudes considered in the main body of the paper. We start from the
factorized $2\to n$ amplitude
\begin{equation}
\label{eq:46}
{\cal A}_{2\to n} = \frac{{\cal A}_{0}}{g^n} \; c_{k_1}\dots
c_{k_n}\;.
\end{equation}
Here ${\cal A}_{0}$ and $g$ are constants, $k_i$ are momenta of particles in
the final state. The cross section of inclusive transition to the 
$n$--particle final states is obtained by integrating over the phase space
volume $\Pi_n$, 
\begin{align}
\notag
\sigma_n(P) &= \int |{\cal A}_{2\to n}|^2\, d\Pi_n(P) \\&=
\frac{|{\cal A}_0|^2}{n!} \int \frac{dk_1 |c_{k_1}|^2}{4\pi g^2
  \omega_{k_1}} \dots \int \frac{dk_n |c_{k_n}|^2}{4\pi g^2
  \omega_{k_n}} \, (2\pi)^2 \delta^{(2)}(k_1 + \dots + k_n - P)\;,
\notag
\end{align}
where we ignored the initial--state factor, introduced the total initial momentum
$P_\mu$ and on--shell frequencies $\omega_{k}^2 = k^2 + m^2$. We use Fourier 
representation of the $\delta$--function and find, 
\begin{equation}
\label{eq:11}
\sigma_n(P) = \frac{|{\cal A}_0|^2}{n!} \int d^2 \lambda \, \left[ f(\lambda)
  \right]^n\mathrm{e}^{iP\cdot \lambda}\;, \qquad \mbox{where} \qquad f(\lambda) = \int
\frac{dk \, |c_k|^2}{4\pi g^2\omega_k}\, \mathrm{e}^{-ik\cdot \lambda} \;.
\end{equation}
If $c_k = c_b$ does not depend on $k$,
$$
f(\lambda) = \frac{|c_b|^2}{2\pi g^2}\,K_0 (m\sqrt{-\lambda^2 +
  i\epsilon \lambda^0})\;. 
$$
Equation~(\ref{eq:11}) shows that $\lambda$ is a typical Compton
wavelength of the final particles, $\lambda\sim k^{-1}$. Summing up the
$n$--particle contributions~(\ref{eq:11}), we obtain
Eq.~(\ref{eq:10}) of Sec.~\ref{sec:grow-ampl-backgr}. 

In Sec.~\ref{sec:pert-expans-at} we consider the factorized
amplitude~(\ref{eq:28}) of the process $2\to n + \Psi_{rt}$. To
simplify summation over the final states in the inclusive cross
section, we relate Eq.~(\ref{eq:28}) to the ordinary $2\to n+m$
amplitudes. To this end we expand the exponent in the final 
state~(\ref{eq:27}) of our amplitude and obtain,
\begin{equation}
\label{eq:45}
{\cal A}_{2\to n + \Psi_{rt}}' =  \sum_{m=0}^{\infty}\,\frac{1}{m!} \,
 \int \frac{dk_{n+1}\, c_{k_{n+1}}^*}{4\pi g \omega_{k_{n+1}}^{(+)}} \; \dots \;
\int \frac{dk_{n+m}\, c_{k_{n+m}}^*}{4\pi g \omega_{k_{n+m}}^{(+)}} \;\; {\cal
A}_{2\to n+m}' \, (2\pi)^2 \delta^{(2)}(P - P_f)\;,
\end{equation}
where $P_f$ is the total momentum of the $(n+m)$--particle final
state. Next, we compare the right--hand side of Eq.~(\ref{eq:28}) with Eq.~\eqref{eq:45}. Substituting $\phi =
\phi_{rt}$ into the wave
functional $\Psi_d[\phi]$ of a coherent state~\cite{Tinyakov:1992dr} with parameters $d_k
\mathrm{e}^{-i\omega_k t_f}$, we find,
\begin{equation}
\label{eq:43}
\Psi_d[\phi_{rt}] \equiv \langle \phi| \Psi_d\rangle =  \exp\left\{\int dk \,  \frac{
  \,d_k \, c_k^*  \, \mathrm{e}^{-ik \cdot
  x_0} }{4\pi  g^2 \omega_k^{(+)}}\right\} \mathrm{e}^{-iE_+ t_0}\,
\Psi_+[\phi_{rt}]\;,
\end{equation}
where we extracted dependence on $x_0$ as explained in
Sec.~\ref{sec:pert-expans-at}, used the form (\ref{eq:34}) of
$\phi_{rt}$, took the limit $t_f \to +\infty 
\cdot\mathrm{e}^{-i\epsilon}$ and introduced wave functional $\Psi_+$ of 
the true vacuum. The final state $\Psi_{rt}$ in
Sec.~\ref{sec:pert-expans-at} has $d_k = c_k$, cf. Eq.~(\ref{eq:27}).
Expanding the exponent in Eq.~(\ref{eq:43}) and 
substituting the series into Eq.~(\ref{eq:28}), we
find, 
\begin{multline}
{\cal A}'_{2\to n+\Psi_{rt}} =  \sum_{m=0}^{\infty}
\frac{1}{m!} \int \frac{dk_{n+1}\, |c_{k_{n+1}}|^2}{4\pi g^2
  \omega_{k_{n+1}}^{(+)}} \; \dots \; \int \frac{dk_{n+m}\, |c_{k_{n+m}}|^2}{4\pi g^2
  \omega_{k_{n+m}}^{(+)}}\, \times \\ \times \, \frac{{\cal A}_0}{g^n} \, c_{k_1} \dots
c_{k_n} \; (2\pi)^2 \, \delta^{(2)}(P -
P_f)\;.
\label{eq:44}
\end{multline}
Here we evaluated the integral over $x_0$ and introduced ${\cal A}_0 =
{\cal A}_{rt} \; \Psi_{+}^*[\phi_{rt}]\;  b_{p_1}^*
b_{p_2}^*/g^2$. Comparing Eqs.~(\ref{eq:45}) and (\ref{eq:44}), one finds
that the amplitudes ${\cal A}_{2\to n+m}'$ have the form
(\ref{eq:46}). We finally obtain the inclusive cross section
(\ref{eq:29}) by summing up the $n$--particle ones in Eq.~(\ref{eq:11}).

To process the dominant amplitude (\ref{eq:33}) of
Sec.~\ref{sec:pert-descr-at}, we do the opposite to the above,
i.e.\ combine the final states into the coherent states $\Psi_d$
with parameters $d_k \equiv c_k \mathrm{e}^{ik\cdot y_0} +
\delta c_k$, where $c_k \mathrm{e}^{ik\cdot y_0}$ are the parameters
of $\Psi_{rt}$ and $\delta c_k$ are arbitrary.  We find,
\begin{align}
\notag
{\cal A}_{2\to \Psi_d} & =   \sum_{n=0}^{\infty} \frac{1}{n!}
\int \frac{dk_1 \; \delta c_{k_1}^*}{4\pi g \,\omega_{k_1}^{(+)}} \, \dots \, \int
\frac{dk_n \; \delta c_{k_n}^*}{4\pi g \,\omega_{k_{n}}^{(+)}} \; {\cal
  A}_{2\to n+\Psi_{f}^{rt}}\\
\label{eq:49}
 & = {\cal A}_{rt} \, \int d^2 x_0 \; \Psi_d^{*}[\phi_{rt}]\int \frac{dk_1
  \, dk_2\, \delta c_{k_1}^* \delta c_{k_2}^*} {16\pi^2 g^2
  \omega_{k_1}^{(+)} \omega_{k_2}^{(+)}} \; \, D_{p_1,\, k_1} \,
D_{p_2,\, k_2} \; \mathrm{e}^{-ix_0 \cdot  (q_1 + q_2)}\;,
\end{align}
In the first line of this expression we expanded $\Psi_d$ in $\delta
c$ using Eq.~(\ref{eq:27}), in the second substituted
Eq.~(\ref{eq:33}) and introduced wave functional $\Psi_d[\phi_{rt}]$,
Eq.~(\ref{eq:43}).

Now, we evaluate the inclusive cross section
\begin{equation}
\label{eq:51}
\sigma(E) = \frac{1}{V^{(2)}} \int {\cal D}d' \, {\cal D} d^* \;
{\cal A}_{2\to \Psi_d} \, {\cal A}_{2\to \Psi_{d'}}^*\,
\mathrm{exp}\left\{-\int \frac{dk \, d_k' \, d_k^*}{4\pi g^2
  \omega_k^{(+)}}\right\}\;, 
\end{equation}
where $V^{(2)}$ is the spacetime volume. First, we note that the
amplitude 
\eqref{eq:51} depends on the arbitrary 
parameter $y_0$. This is the freedom of choosing the background for
expansion, it disappears after resummation of perturbative series. We
fix the freedom requiring
\begin{equation} 
y_0 = x_0' \;, \qquad\qquad y_0' = x_0\;,
\end{equation}
where $y_0'$ and $x_0'$ come from the complex conjugate
amplitude~(\ref{eq:49}) in the integral for the inclusive cross section. 
Second, the parameters $\delta c_k^*$ in the prefactor can be obtained by
varying the exponent $\exp\{\int dk \, \alpha_k \delta c_k^* / (4\pi
g^2\omega_k^{(+)})\}$ with respect to $\alpha$ at $\alpha=0$. The
integral in Eq.~(\ref{eq:51}) therefore can be evaluated using the
functional  
\begin{align}
\Pi [\alpha\, , \alpha'^*] & = \int {\cal D} \delta c' \, {\cal D} \delta
c^* \; \Psi_d^{*}[\phi_{rt}]\; \Psi_{d'}[\phi_{rt}] \;
\mathrm{exp}\left[\int \frac{dk}{4\pi g^2 
  \omega_k^{(+)}} (\alpha_k \delta c_k^*  + \alpha_k'^* \delta c_k' -
d_k' \, d_k^*)\right]\notag\\
& = \mathrm{e}^{-iE_+ \lambda^0} \, \left|\Psi_+[\phi_{rt}]\right|^2 \,
\exp \left[ \int \frac{dk}{4\pi g^2 \omega_k^{(+)}} (\alpha_k
\alpha_k'^* + |c_k|^2 \mathrm{e}^{-ik\cdot \lambda})\right]\;,
\label{eq:52}
\end{align}
where we denoted $d_k' \equiv c_k\mathrm{e}^{ik\cdot y_0'} + \delta c_k'$,
substituted Eq.~(\ref{eq:43}), evaluated the integrals over $\delta c^*$,
$\delta c'$ and denoted $\lambda \equiv y_0 - x_0 = x_0' - x_0$. We
see that variations over $\alpha_k$ and $\alpha_k'^{*}$ give
contraction  rule
for the final--state variables
$\delta\mbox{\hspace{-0.8mm}}\overbracket[0.5pt]{c_{k_1} \,    \delta}
\mbox{\hspace{-0.8mm}}c_{k_2}^* = 4\pi g^2 \omega_k^{(+)} 
\delta (k_1 - k_2)$.

Substituting Eqs.~(\ref{eq:49}) and (\ref{eq:52}) into
Eq.~(\ref{eq:51}), one obtains the inclusive cross section
(\ref{eq:53}) containing the integral over $\lambda = x_0' - x_0$;
integration with respect to $(x_0 + x_0')/2$ gives  the two--volume
$V^{(2)}$ which is canceled in Eq.~(\ref{eq:51}). 

\section{High--frequency tail of the semiclassical solution}
\label{sec:semicl-solut-at}
Let us evaluate high--frequency asymptotics of the saddle--point
solution $\phi_s(x)$. To this end we represent $\phi_s(x) = \phi_0(x)
+ \delta \phi(x)$ as a sum of soft nonlinear background $\phi_0$
and high--frequency part $\delta\phi \ll \phi_0$ evolving on top of it. Functions
$\phi_0$ and $\delta \phi$ contain modes with $k<\Lambda$ and
$k>\Lambda$, respectively. One rewrites the field equation as 
\begin{equation}
\label{eq:19}
\Box \delta \phi(x) = J(x)\;,
\end{equation}
where $J$ is a contribution of $\phi_0$ and $\delta \phi$ is
ignored in potential terms. We solve Eq.~(\ref{eq:19}),
\begin{equation}
\label{eq:20}
\delta \phi(x) = -\int \frac{d^2k}{(2\pi)^2}\; \frac{J(k)\,
  \mathrm{e}^{-ik\cdot x}}{k^2 - i\epsilon
  k^0}  +  \int\limits_{k>\Lambda} \frac{dk}{4\pi \omega_k}
\left[c_k \mathrm{e}^{-ik\cdot x}  + c_k^* \mathrm{e}^{ik\cdot
    x}\right]\;,
\end{equation}
using the two--dimensional Fourier image of the source $J(k)$ and
arbitrary on--shell waves in the second term with $k^\mu = (\omega_k,\, k)$ and $\omega_k
= |k|$. The solution (\ref{eq:20}) is real as $t\to +\infty$ in
accordance with 
Eq.~(\ref{eq:57}). In the infinite past, i.e.\ $t\to iT - \infty$,
Eq.~(\ref{eq:20}) takes  the form (\ref{eq:17}) with
\begin{equation}
\label{eq:22}
a_k = c_k + J_k^{-}\;, \qquad b_k^* = c_k^* + J_k^+\;, \qquad \mbox{and}
\qquad J_k^{\pm} \equiv \pm iJ(\mp k)\Big|_{k^0 = \omega_k} \;.
\end{equation}
We finally solve the initial condition \eqref{eq:39} and obtain,
\begin{equation}
\label{eq:21}
c_k = \frac{\gamma_k (J_k^+)^* - J_k^-}{ 1 - \gamma_k}\;, \qquad \qquad
\gamma_k \equiv \mathrm{e}^{-2\omega_k T - \theta}\;. 
\end{equation}
\begin{wrapfigure}[12]{r}{4.9cm}

\hspace{3mm}\includegraphics[width=4.4cm]{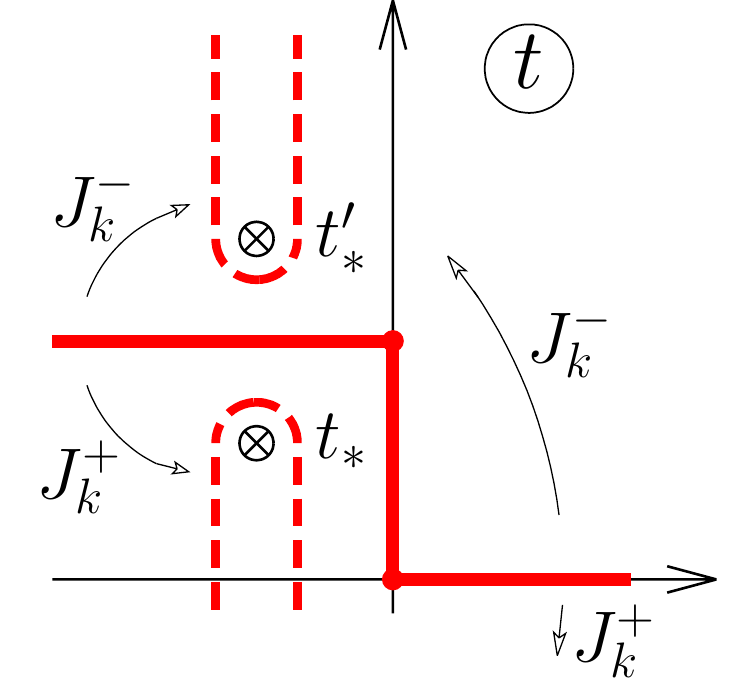}
\vspace{-2mm}
\caption{Contours for $J_k^\pm$.\label{fig:contourJ}}
\end{wrapfigure}
Note that Eq.~(\ref{eq:19}) and its solution
(\ref{eq:20}),~(\ref{eq:21}) are valid only at $k>\Lambda \gg m$.

One notices that the sources $J_k^{\pm}$ are exponentially
sensitive to $k$. Indeed, the $t$--contours in the Fourier transforms
\begin{equation}
J_k^{\pm} = \pm i\int d^2 x \, J(x)\,
\mathrm{e}^{\mp ik\cdot x}\;\Big|_{k^0 = \omega_k}
\end{equation}
can be deformed into the upper and lower parts of the complex time plane
until they hit the singularities $t_*$ and $t_*'$ of the
solution\footnote{Recall that $J(x)$ 
  is related to $\phi_0(x)$.}, see Fig.~\ref{fig:contourJ}.
  At large
$k$ the integrals receive the dominant contribution near the 
singularities, and
\begin{equation}
\label{eq:23}
J_k^{+} \sim J_0^+ \; \mathrm{e}^{-i\omega_k t_*}\;, \qquad \qquad
J_k^{-} \sim J_0^- \; \mathrm{e}^{i\omega_k
  t_*'}
\end{equation}
where $J_0^{\pm}$ are constants.

We finally compute the energy of the initial particles with momentum $k$
in Eq.~(\ref{eq:14}) using Eqs.~(\ref{eq:22}), (\ref{eq:21}), 
\begin{equation}
\label{eq:25}
\varepsilon_k \equiv \frac{a_k b_k^*}{4\pi} = \frac{|(J_k^+)^* -
  J_k^-|^2}{16\pi\sinh^2(\omega_k T + \theta/2)}\;,
\end{equation}
Estimating the largest term in the nominator by Eq.~(\ref{eq:23}),
one obtains the high--frequency asymptotic (\ref{eq:18}) with $T_* =  T  +
\min(-\mathrm{Im}\, t_*, \, \mathrm{Im}\,
t_*')$. Equation~(\ref{eq:25}) obeys\footnote{In this case the soft
  background $\phi_0$ and related sources $J_k^{\pm}$ are almost
  independent of $T$.} the rescaling
property (\ref{eq:24}) of Sec.~\ref{sec:transitions-at-ee_rt}.  


\end{document}